**PHYSICS**

# New constraints on axion-like dark matter using a Floquet quantum detector

Itay M. Bloch[1,2]†, Gil Ronen[2,3]†, Roy Shaham[2,4], Ori Katz[3], Tomer Volansky[1], Or Katz[4]*‡

Dark matter is one of the greatest mysteries in physics. It interacts via gravity and composes most of our universe, but its elementary composition is unknown. We search for nongravitational interactions of axion-like dark matter with atomic spins using a precision quantum detector. The detector is composed of spin-polarized xenon gas that can coherently interact with a background dark matter field as it traverses through the galactic dark matter halo. Conducting a 5-month-long search, we report on the first results of the Noble and Alkali Spin Detectors for Ultralight Coherent darK matter (NASDUCK) collaboration. We limit ALP-neutron interactions in the mass range of $4 \times 10^{-15}$ to $4 \times 10^{-12}$ eV/$c^2$ and improve upon previous terrestrial bounds by up to 1000-fold for masses above $4 \times 10^{-13}$ eV/$c^2$. We also set bounds on pseudoscalar dark matter models with quadratic coupling.

## INTRODUCTION

A plethora of multiscale astrophysical and cosmological evidence suggests that roughly 85% of the matter in our universe is unlike the matter we see around us. Despite almost a century of research, the evidence for this so-called dark matter (DM) is purely gravitational. As a result, its entire particle identity including its mass, spin, and interactions with itself and with other particles remains unknown.

Various theoretical models propose an abundance of candidates that would explain the physical nature of DM. A well-motivated one is a postulated particle named the axion (1, 2), originally introduced to solve the strong CP problem (3). Over the years, many generalizations to the axion have been postulated, and they are collectively known as axion-like particles (ALPs) (4). These ALPs can be produced in the early universe and can account for the observed phenomena associated with DM. While the uncertainty for the ALP mass spans many orders of magnitudes, a particularly interesting range is that of ultralight masses (5). In this regime, the ALP De-Broglie wavelength is considerably longer than the length of the detector, and in addition, the number of particles within a single De-Broglie wavelength cubed (roughly the classical volume a single particle occupies) is much larger than 1. This implies that any interaction of ultralight ALPs with other particles such as protons, neutrons, electrons, and photons would be coherently enhanced and more easily detected (6–14). Moreover, the coupling between ultralight ALP DM with electron and nuclear spins can be manifested in the form of anomalous magnetic fields that induce an oscillatory energy shift at a characteristic frequency that depends on the ALP mass (15).

Various groups search for cosmological DM using astronomical observations and terrestrial detectors. Comagnetometers and nuclear magnetic resonance (NMR) sensors, in particular, are compact sensors that feature enhanced sensitivity to regular and anomalous magnetic fields. These sensors are composed of a dense ensemble of spin-polarized nuclei in a gaseous, liquid, or solid phase, whose collective response to regular or anomalous fields is measured by a precision magnetometer (6, 9–13, 16–25).

While these sensors have long been applied in various disciplines including medicine (26), chemistry (27, 28), geology (29), physics, and engineering (30), their application for the search of cosmological DM is only at its infancy, with potential unprecedented sensitivities (31–33). Recently, these sensors have set new terrestrial constraints on the coupling of neutrons to ALP DM with masses $m_{DM} \lesssim 4 \times 10^{-13}$ eV/$c^2$ (6) by using an in situ atomic magnetometer. However, this in situ magnetometry is typically limited to measurements of small ALP masses, the reason for which can be traced back to the large difference between the gyromagnetic ratios of the nuclear spins and the electronic spins that comprise the two magnetometers.

Application of strong time-modulated fields, and Floquet engineering methods in particular, provides exquisite control over properties of materials. These methods have been widely applied in various disciplines in condensed matter and atomic physics, enabling control over the topology and band structure of materials (34, 35), the formation of time crystals (36), and modification of the effective gyromagnetic ratio of atoms and their response to external fields (37, 38). Utilization of Floquet fields have long enabled to enhance the sensitivity of NMR sensors at high frequencies (39), and recently, it was proposed as an eminent avenue to enhance the performance of DM field detectors (40). However, constraints on the coupling of DM (and in particular ALPs) to fermions using Floquet techniques have never been realized until this work.

Here, we report on new experimental constraints on the ALP-DM interactions with neutrons. The results, first from the NASDUCK collaboration (Noble and Alkali Spin Detectors for Ultralight Coherent darK matter), rely on measurements that took place over a period of 5 months using a dense spin-polarized ensemble of $^{129}$Xe atoms, whose response to anomalous fields was measured using an in situ precision rubidium Floquet magnetometer (see Fig. 1). The presence of the Floquet field enabled us to expand our search by more than an order of magnitude in masses, placing strong constraints in the mass range $4 \times 10^{-15}$ eV/$c^2 < m_{DM} < 4 \times 10^{-12}$ eV/$c^2$. We improve on the current terrestrial limits on the coupling to neutrons by as much as three orders of magnitude. We also cast bounds on quadratic interactions (13), improving all existing bounds for some masses within the range of $2 \times 10^{-14}$ eV/$c^2 < m_{DM} < 7 \times 10^{-13}$ eV/$c^2$ for neutron-DM quadratic interactions. Last, we also cast additional model-dependent bounds on the coupling of ALPs to protons and discuss their model uncertainty.

[1]School of Physics and Astronomy, Tel-Aviv University, Tel-Aviv 69978, Israel. [2]Rafael Ltd., IL-31021 Haifa, Israel. [3]Department of Applied Physics, Hebrew University of Jerusalem, 9190401 Jerusalem, Israel. [4]Department of Physics of Complex Systems, Weizmann Institute of Science, Rehovot 76100, Israel.
*Corresponding author. Email: or.katz@duke.edu
†These authors contributed equally to this work.
‡Present address: Department of Electrical and Computer Engineering, Duke University, Durham, NC 27708, USA.

## RESULTS

### Interaction of ALP DM with spins

ALPs are pseudoscalars that can couple to neutrons in the form of an oscillatory magnetic-like field. The field amplitude $b_{DM}$ is related to the DM local energy density $\rho_{DM}$, whereas its frequency $f_{DM}$ is related to the ALP DM mass $m_{DM}$. The coupling of ALPs to spins is described by the interaction Hamiltonian

$$H = \gamma_{Xe} \langle \cos\left(2\pi f_{DM} t + \phi_{DM}\right) \mathbf{b}_{DM} \rangle_{\mathbf{v}, \phi_{DM}} \cdot \mathbf{I}_{Xe} \quad (1)$$

where $\mathbf{I}_{Xe}$ denotes the nuclear spin-1/2 operator of xenon and $\gamma_{Xe}$ denotes its gyromagnetic ratio. The DM field oscillates at frequency $f_{DM} = m_{DM}(c^2 + \mathbf{v}^2/2)/h$ (and with a random initial phase $\phi_{DM}$), with $\mathbf{v}$ being its velocity. The DM's wave function is a narrow wave packet in the frequency domain with a center slightly above $m_{DM}c^2/h$ and a spread (as well as the offset from $m_{DM}c^2/h$) that depends on the velocity distribution of the DM, which is moving in the galaxy with respect to Earth [see (41) for further details]. According to the standard halo model, the mean and the SD of $\mathbf{v}$ are both of order the virial velocity $v_{vir} \approx 220$ km/s (42), and as a result, the field remains coherent for a considerably long time $h/(m_{DM}v_{vir}^2)$, which corresponds to about $2 \times 10^6$ oscillations. In Eq. 1, $\langle \cdot \rangle_{\mathbf{v}, \phi_{DM}}$ denotes the averaging over the velocity and phase distributions [see (41) for further details], while $\mathbf{b}_{DM}$ is the anomalous magnetic field that is defined by

$$\mathbf{b}_{DM} = \epsilon_N g_{aNN} \sqrt{2\rho_{DM}\hbar c^3} \, \mathbf{v}/\gamma_{Xe} \quad (2)$$

Here, $g_{aNN}$ is the ALP-neutron coupling coefficient and $\rho_{DM} = 0.4$ GeV/$(c^2$cm$^3)$ (42). $\epsilon_N$ is the fractional contribution of neutrons to the nuclear spin. Because the $^{129}$Xe nucleus has a valence neutron, the model uncertainty of $\epsilon_N$ is relatively small, and contribution from coupling to its protons is about two orders of magnitude smaller. We adapt $\epsilon_N = 0.63$, corresponding to the smallest estimation in (43), using the $\Delta$NNLO$_{GO}$(394) model of the nuclear interactions. $\mathbf{b}_{DM}$ is considered an anomalous magnetic field because the Hamiltonian in Eq. 1 is independent of the gyromagnetic ratio (note that $\mathbf{b}_{DM} \propto 1/\gamma_{Xe}$), which is associated with the spin coupling of fermions to regular magnetic fields and is thus unrelated to the interaction with ALPs.

In addition to ALP models, we also cast bounds on a pseudoscalar DM model, which interacts quadratically with neutrons (13). In this model, the anomalous field is given by

$$\mathbf{b}_{DM} = \epsilon_N g_{N-Quad}^2 \frac{2\rho_{DM}\hbar^2 c^2}{m_{DM}} \mathbf{v}/\gamma_{Xe} \quad (3)$$

Here, the DM oscillates at $f_{DM}^{Quad}(m_{DM}) = f_{DM}^{ALP}(2\,m_{DM})$, implying that bounds on quadratic-type interactions are derived in the mass range $2 \times 10^{-15}$ eV/$c^2 < m_{DM} < 2 \times 10^{-12}$ eV/$c^2$.

### Experimental setup and detection mechanism

The heart of our sensor consists of spin-polarized $^{129}$Xe atoms, whose collective response to time-varying fields is detected via an in situ precision optical magnetometer made of rubidium vapor. The atoms are encapsulated in a small cubical glass cell that is maintained at 150°C and surrounded by magnetic shields as shown in Fig. 1. We continuously polarize the Rb spins at their electronic ground state via optical pumping. The Rb polarizes, in turn, the nuclear spins of $^{129}$Xe via spin-exchange collisions along the $\hat{z}$ axis, effectively maintaining $\sim 3 \times 10^{16}$ fully spin-polarized nuclei. Further details on the experimental configuration are described in Materials and Methods and (*41*).

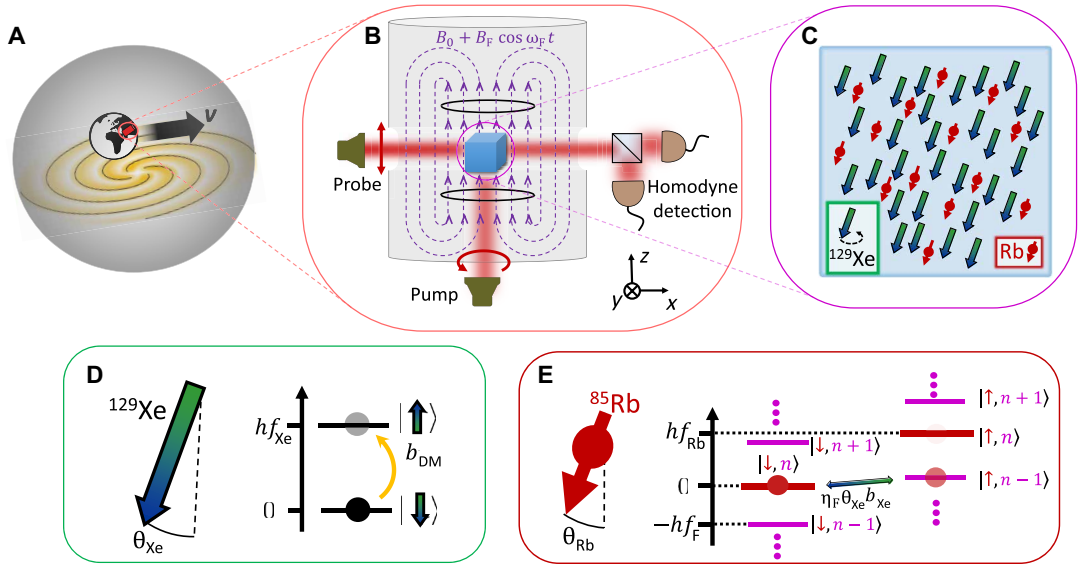

**Fig. 1. Floquet quantum detector for the search of ultralight axion-like DM. (A)** As Earth moves across the Milky Way galaxy, it traverses the DM halo with a mean virial velocity $v_{vir}$. **(B and C)** The Floquet detector is composed of a dense ensemble of spin-polarized $^{129}$Xe gas, which can resonantly interact with the moving axion-like DM. The interaction is in the form of an anomalous magnetic field, penetrating the detector shields that deflect regular magnetic fields. The spin precession is monitored via an in situ optical magnetometer using $^{85}$Rb vapor that is magnetically driven by a strong Floquet field $B_F$. **(D)** Energy-level structure of the nuclear spin of $^{129}$Xe. The DM field oscillating near the NMR resonance frequency of the xenon with amplitude $b_{DM}$ can drive collective spin flips of the ensemble in a coherent manner, rotating the net direction of the spin-polarized ensemble at an angle $\theta_{Xe}$. **(E)** Floquet spectrum of the $^{85}$Rb spins dressed by $n$ RF photons. Collective spin flips of the polarized $^{85}$Rb ensemble by the slowly precessing xenon field ($\theta_{Xe}b_{Xe}$) are greatly enhanced when the energy splitting of the Rb is large ($f_{Rb} \gtrsim \Gamma_{Rb}$). For example, absorption of an RF photon of the Floquet field in the transition ($|\downarrow, n\rangle \to |\uparrow, n-1\rangle$) is enhanced by a factor $\eta_F$ compared to a spin flip ($|\downarrow\rangle \to |\uparrow\rangle$) in the absence of the Floquet drive ($n = 0$). This transition bridges between the large frequency mismatch of the electron ($^{85}$Rb) and nuclear ($^{129}$Xe) spin resonances and enables efficient detection at frequencies higher than previously measured.

An anomalous DM field pointing in the $xy$ plane of the detector and interacting with the $^{129}$Xe nucleons would collectively tilt them off the $\hat{z}$ axis by an angle $\theta_{Xe}$ and force their precession at a frequency $f_{DM}$. In the presence of an axial magnetic field, this tilt is suppressed outside of a narrow frequency band with width $\Gamma_{Xe}$ centered around the NMR frequency $f_{Xe}$ (corresponding to the precession frequency around the axial field) [We note that, in our setup, $f_{Xe} = \gamma_{Xe}(B^z_{ext} + b_{Rb})$, where $B^z_{ext}$ is the external field in the $z$ direction and $b_{Rb}$ is the effective magnetic field induced by the Rb atoms via the spin-exchange interactions]. Thus, for $f = f_{DM}$

$$\theta_{Xe}(f = f_{DM}) = \frac{\gamma_{Xe}\, b_{DM}}{2 \,|\, i\,\Gamma_{Xe} + f_{Xe} - f_{DM}\,|} \tag{4}$$

Here, $\gamma_{Xe} = -1.18$ kHz/G is the gyromagnetic ratio of xenon, and $\Gamma_{Xe} \approx 0.3$ Hz is the measured decoherence rate of the xenon spins.

From Eq. 4, we thus learn that the noble-gas spins efficiently respond to the anomalous ALP field if it oscillates at a frequency $f_{DM}$ that resonates with the NMR frequency $f_{Xe}$. Our setup is capable of efficiently sensing $f_{DM}$ in the 1- to 1000-Hz range.

To measure the precession of the noble-gas spins, we use the rubidium as an optical magnetometer. Using a linearly polarized optical probe beam, we measure the collective spin of the rubidium along the $\hat{x}$ axis via its imprint on the polarization of the probe beam, which rotates after traversing the alkali medium. The polarization rotation is subsequently measured with a set of differential photodiodes in a homodyne configuration (*44*–*46*). While the polarized rubidium spins are initially oriented along the $\hat{z}$ axis, they are tilted by an angle

$$\theta_{Rb}(f = f_{DM}) = \frac{\gamma_{Rb}\, b_{Xe}\, \theta_{Xe}}{|\, i\,\Gamma_{Rb} + f_{Rb} - f_{DM}\,|} \tag{5}$$

Here, $b_{Xe}\theta_{Xe}$ is the transverse spin-exchange field. $b_{Xe}$ is proportional to the magnetic field produced by the spin-polarized xenon atoms and is enhanced by a large factor of $\kappa_0 = 518$ owing to the Fermi contact interaction with the rubidium (*47*). We stress that this factor is a virtue of having an in situ magnetometer and allows for an improved sensitivity of the detector. $f_{Rb}$ is the electron paramagnetic resonance (EPR) frequency (analogous to the NMR frequency for nuclear spins) of $^{85}$Rb spins, $\gamma_{Rb} = 467$ kHz/G is the gyromagnetic ratio of the $^{85}$Rb isotope (72% abundance), and $\Gamma_{Rb} = 6.6$ kHz is their decoherence rate.

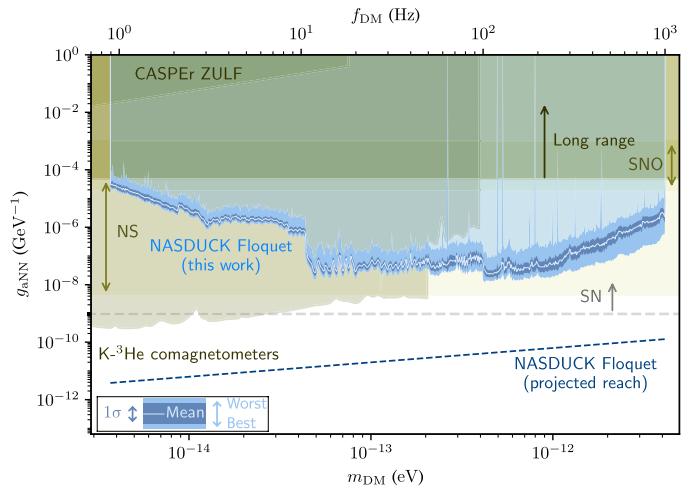

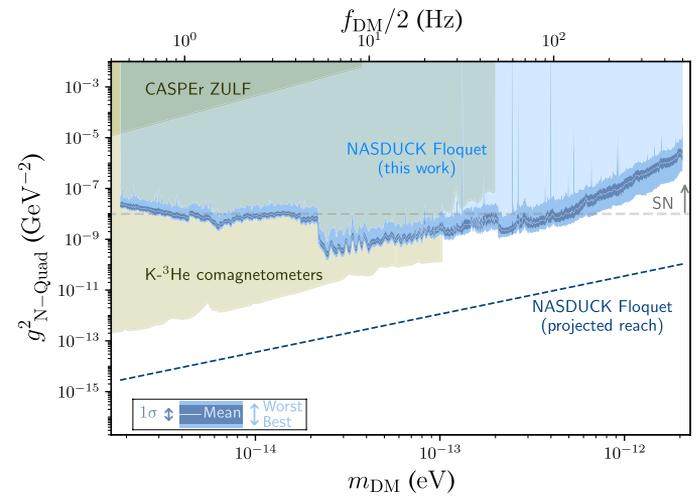

**Fig. 2. Constraints on ALP-neutron couplings.** In this work, we derive the constraints using the Floquet quantum detector by the NASDUCK collaboration as a function of the ALP mass. All presented constraints calculated in this paper correspond to 95% C.L. of the bound. Because of the finite resolution of the figure and given the dense set of measurements, the limits appear as a bright blue band. The width of this band denotes the strongest and weakest values around each mass point. The precise tabulated bounds can be found in (*48*). The bright blue solid line shows a binned average of the bound, while its 1σ variation is shown in dark blue band (both calculated in log space at a binning resolution of 1% of the mass). The light transparent blue region shows the exclusion region for the ALP-neutron couplings. The dashed dark blue line shows the projected sensitivity of this experiment, as discussed in the text. The olive green regions show other terrestrial constraints, including the CASPEr ZULF experiments (*12*, *49*), K-$^3$He comagnetometer bounds (*6*, *18*, *50*), and long-range constraints on ALP-neutron (*18*) couplings. In beige, the agreed astrophysical excluded region from solar ALPs unobserved in the SNO (*51*) and from NS cooling (*52*, *66*–*68*) is shown. The region above the gray dashed line is excluded by SN cooling considerations and neutrino flux measurements (*4*, *53*, *54*). The SN cooling constraint strongly relies on the unknown collapse mechanism, and hence, the limits should be taken with a grain of salt (*55*).

**Fig. 3. Constraints on neutron-DM couplings of quadratic type.** The quadratic constraints (*13*) are derived as a function of the ALP mass, using the Floquet quantum detector by the NASDUCK collaboration in this work. All presented constraints calculated in this paper correspond to 95% C.L. of the bound. As in Fig. 2, because of the finite resolution of the figure and given the dense set of measurements, the limits appear as a bright blue band. The width of this band denotes the strongest and weakest values around each mass point. The precise tabulated bounds can be found in (*48*). The bright blue solid line shows a binned average of the bound, while its 1σ variation is shown in dark blue band (both calculated in log space at a binning resolution of 1% of the mass). The light transparent blue region shows the exclusion region for this model. The dashed dark blue line shows the projected sensitivity of this experiment, as discussed in the text. The olive green regions show other terrestrial constraints, including the CASPEr ZULF experiments (*12*, *49*) and the recasted K-$^3$He comagnetometer bounds (*6*, *18*, *50*). The region above the dashed gray line is excluded by SN cooling considerations (*13*). We stress that the SN cooling constraint strongly relies on the unknown collapse mechanism, and hence, the limits should be taken with a grain of salt (*55*).

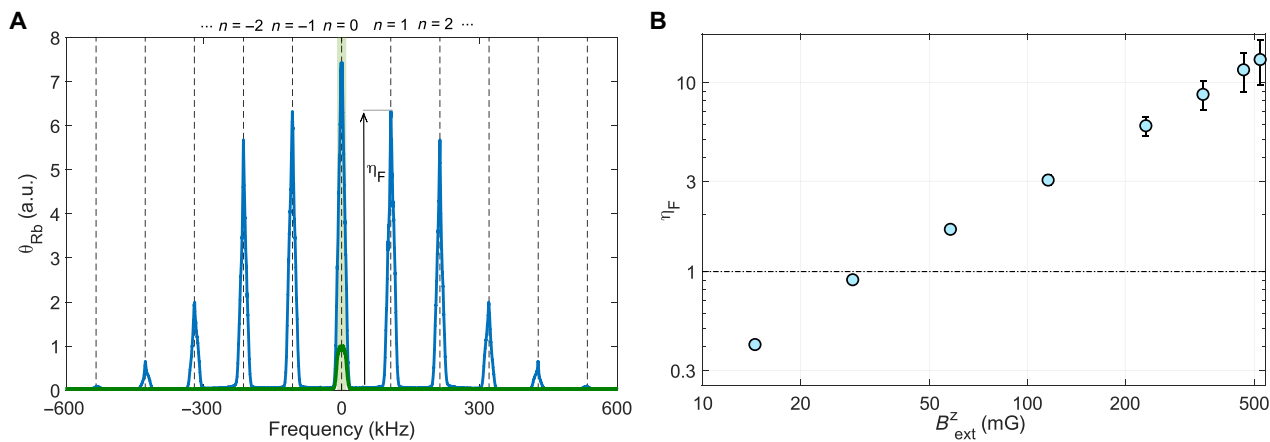

**Fig. 4. The Floquet modulation and the corresponding effective enhancement of the bandwidth.** (**A**) The response (root mean square) in arbitrary units (a.u.) of the optical (Rb) magnetometer (green line) to an input low-frequency transverse magnetic noise in the absence of the Floquet field. The injected noise spectral content is cut off at 10 kHz (light green shade), and the axial magnetic field is taken to be $B_{ext}^z = 0.23$ G, corresponding to a resonance frequency of $f_{Rb} = 105$ kHz. In the absence of the Floquet modulation, the response has the same spectral support as the input field and its peak is normalized to unity. In the presence of a strong axial Floquet field, the magnetometer response (blue line) is altered substantially, mapping and enhancing the input field at integer multiples of the Floquet frequency (indicated with dashed lines). $\eta_F$ denotes the enhancement of the magnetic response near the first Floquet band over the maximal response in the absence of the Floquet field. (**B**) Measured enhancement bandwidth factor, $\eta_F$, as a function of the axial magnetic field. The Flouqet modulation, which is tuned to the Rb resonance frequency, is shown to improve the magnetometer response over the unmodulated case and by that enable recovering of much of its maximal sensitivity (dashed line).

The angle $\theta_{Rb}$ that is measured by the optical magnetometer is proportional to the DM anomalous field $b_{DM}$. To detect this field efficiently, Eqs. 4 and 5 indicate that both the NMR and EPR frequencies should be brought in resonance with the anomalous DM field, satisfying $|f_{Xe} - f_{DM}| \lesssim \Gamma_{Xe}$ and $|f_{Rb} - f_{DM}| \lesssim a_{Rb}$ simultaneously. While both NMR and EPR frequencies depend linearly on the magnetic field $B_{ext}^{\ z}$ and have small nonzero offsets, their substantially different slopes $\gamma_{Xe}$ and $|\gamma_{Rb}| \approx |400\gamma_{Xe}|$ hinder simultaneous resonance for high frequencies, where $f_{DM} \gg (\gamma_{Xe}/\gamma_{Rb})\Gamma_{Rb}$. Thus, in this regime, while the NMR condition can be satisfied, the alkali spins remain out of resonance and the detector's sensitivity is strongly impeded.

To address this problem and increase the range of masses in which DM can be detected, we apply an additional strong Floquet field that enhances the response of the detector and bridges the NMR and EPR frequency gap. The application of the Floquet field enables one to recover the sensor sensitivity at high axial fields, bridging the resonance mismatch between the bare magnetic resonances of the nuclear and alkali spins. Characterization of the Floquet operation and characterization of the detector's background and performance are detailed in Materials and Methods.

### The search
#### Data acquisition
We search for anomalous magnetic fields oscillating in the range of $f_{DM} = 1$ to 1000 Hz (corresponding to $m_{DM}$ in the range of $4 \times 10^{-15}$ to $4 \times 10^{-12}$ eV/$c^2$). To maximize the realized sensitivity, for each searched frequency, we tuned the axial magnetic field to bring the NMR frequency $f_{Xe}(B_{ext}^z)$ near resonance with $f_{DM}$ and typically recorded the sensor response for $(2 - 20) \times 10^5$ oscillations. To cover all frequencies within the search range, we scanned the magnetic field in fine steps of about 0.2 mG to maintain overlap in the NMR frequencies of neighboring measurements. We have scanned the entire range of frequencies several times by slightly shifting the NMR frequencies to get ample measurements of any given frequency at different sensitivities. As a result, the search consisted of almost 3000 measurements, taken during a period of 5 months. To characterize the sensor response and validate its stability during the long search period, each measurement was preceded and followed by a set of calibration measurements. The calibration measurements automatically tuned the parameters of the oscillatory Floquet field and characterized the response function of the sensor. Data processing is described in Materials and Methods, and detailed search and calibration protocols are given in the Supplementary Materials (*41*).

#### Detection capability
As DM ALP signals feature extremely long coherence time compared to the sensor backgrounds, they can potentially be directly detected and distinguished from the background. In the frequency domain, a signal centered at $f_{DM}$ would have an ultranarrow bandwidth with a quality factor of about $\sim 2 \times 10^6$, thus being distinguishable from the noise that is approximately white within that bandwidth [see (*41*)]. Sideband analysis further enables to differentiate between the signal and background. We use measurements at magnetic fields in which the NMR frequency is off-resonant and the sensitivity to anomalous fields is negligible as control measurements. These measurements enable us to identify frequencies in which the background has coherent properties that require specialized analysis procedures that rely on the different responses instead of the coherence times of signal and background. Last, multiple repetitions of measurements enable to exclude transient noise and differentiate it from the coherent ALP signal.

#### Search results
We use the log-likelihood ratio test to constrain the presence of ALP DM with 95% confidence level (C.L.) bounds, presented on ALP-neutron couplings in Fig. 2. The constraints cover the entire mass range between $4 \times 10^{-15}$ and $4 \times 10^{-12}$ eV/$c^2$, measured with high resolution roughly set by the reciprocal measurement time. The resulting limits are presented by the bright blue band and are tabulated in (48). Because of the finite resolution of the figures, the limits appear simply as a bright blue band whose width reflects the strongest and weakest values around each frequency point. To convey the typical bound, we further present a binned average of the bound (bright blue solid line) and its 1σ variation (dark blue band), both calculated in log space at a binning resolution of 1% of the mass. Notably, all values of the ALP-neutron couplings above the bright blue band are excluded (light transparent blue region). The olive green regions show other terrestrial constraints, including the CASPEr ZULF experiments (12, 49), K-$^3$He comagnetometer bounds (6, 18, 50), and long-range constraints on ALP-neutron (18) couplings. In beige, the astrophysical limits from not observing solar ALPs in the Solar Neutrino Observatory (SNO) (51) and neutron star (NS) cooling considerations from ALP-neutron interactions (52) are shown. The region above the dashed gray line is excluded by supernova (SN) cooling considerations and neutrino flux measurements (4, 53, 54).

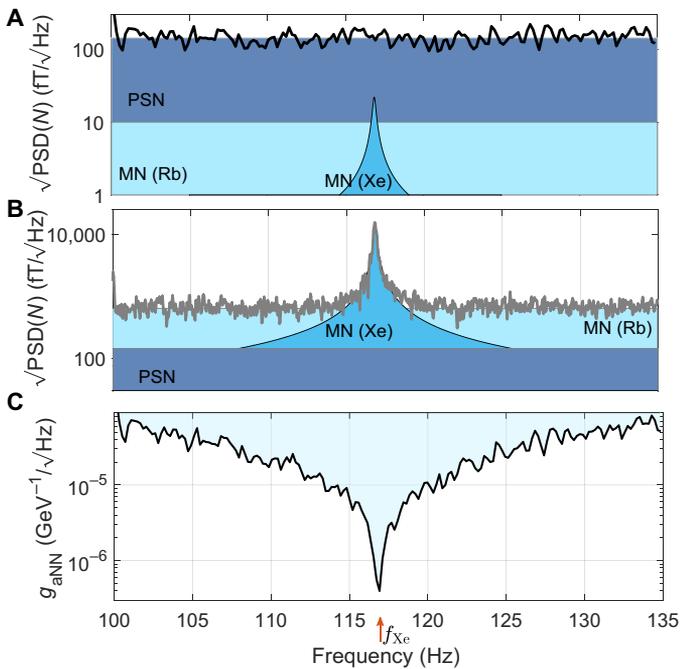

**Fig. 5. Noise spectral density measurement and its implication on the detector sensitivity.** (**A**) Measured noise spectral density PSD(N) (black line) near the xenon resonance frequency in the presence of a Floquet field and an axial field $B_{ext}^z = 0.1$ G. The shaded regions denote the different estimated noise contributions of optical photon shot noise (PSN; dark blue) and magnetic noise (MN) coupled to Rb spins (light blue) and Xe spins (blue). (**B**) Sensor response (gray line) to an injected, white magnetic noise $\delta B_x = 5.6\,\mathrm{pT}/\sqrt{\mathrm{Hz}}$. By tuning the magnetometer parameters to be sensitive predominantly to one direction in the $xy$ plane, one is able to disentangle the response of the xenon spins from the injected noise. The figure shows the resonant behavior of the xenon in the presence of a dominant magnetic noise. Shaded regions are the same as the top figure. (**C**) Sensitivity to anomalous fields as inferred from the calibrated response of the magnetometer and the noise characteristics in (A), assuming coherent and deterministic ALPs (black line). When the photon shot noise (flat) spectrum is dominant, the signal and hence the sensitivity are enhanced near the xenon resonance frequency (red arrow).

Figure 3 presents with constraints on neutron-DM couplings of quadratic type. The new bounds of the NASDUCK Floquet detector use the same coloring conventions as the bounds of the ALP-neutron interactions. Similarly, the olive green regions show other terrestrial constraints, from the CASPEr ZULF experiment (12, 49), and bounds converted in (6), using data from (18, 50) from K-$^3$He comagnetometers. The regions above the dashed gray lines are excluded by the more uncertain [see (55) for further details] bounds that arise from SN cooling considerations (13).

Current models of the nuclear structure of $^{129}$Xe predict that its spin composition has a nonzero fractional contribution from protons $\epsilon_P$. If $\epsilon_P$ is nonzero, then the search data can also be used to cast additional bounds on the coupling strength of ALPs with protons by carrying the substitution $\epsilon_N g_{aNN} \rightarrow \epsilon_P g_{aPP}$ in Eq. 2 (assuming that $g_{aNN} = 0$ for this particular likelihood test). In (41), we present the bounds obtained by our data for four different values of $\epsilon_P$ associated with different nuclear structure models (43, 56). For these values, the new model-dependent bounds we derive improve other terrestrial bounds on the coupling of ALPs with protons. However, we strictly emphasize that the reliability of these model-dependent proton coupling bounds should be taken with a grain of salt. The uncertainties of the nuclear structure models could be quite large as they are not sufficiently quantified for $^{129}$Xe (57), and possibly $\epsilon_P$ could even vanish.

Estimation of future performance for casting stronger bounds than our current detector is presented by the dashed dark blue line based on usage of low-noise ferrite shields (58) and multiple passages of the probe beam within the cells (59), expecting a white noise floor of $1\,\mathrm{fT}/\sqrt{\mathrm{Hz}}$. The reach is estimated for measurements of $2 \times 10^6$ oscillations for each frequency, corresponding to about 2 years of measurement using a single detector.

Our newly derived limits on the ALP-neutron coupling substantially improve the existing terrestrial limits in the mass range of $2 \times 10^{-13}$ to $4 \times 10^{-12}$ eV/$c^2$, complementing the yet stronger astrophysical constraints. Notably, astrophysical constraints typically suffer from substantial systematic uncertainties that render them less certain.

## DISCUSSION
In summary, we presented new constraints on the ALP DM couplings to neutrons, substantially improving previous bounds. Our detector used dense xenon spins and in situ Floquet magnetometer, which enabled the extension of ALP masses to higher values. In addition, new bounds on neutron-quadratic type interactions were also cast.

NMR detectors are frontier technology to search for new physics, which, for ALPs, can potentially reach the limits of QCD axions in the standard quantum limit (33). Practically, however, these sensors are often limited by the magnetic noise floor and the realized detector sensitivity that nonetheless are expected to exceed astrophysical bounds for coupling with neutron spins. In addition, similar searches using nuclei whose proton spin component is known with higher certainty could extend such searches and cast reliable bounds on the coupling of ALPs with protons.

## Note added
Toward the end of our search and during the late stages of the analysis of the recorded data, we became aware of the study of Jiang *et al.* (60), which uses an NMR detector to search for nucleon-ALP coupling. Jiang *et al*.'s (*60*) study finds similar sensitivity, but it does not use Floquet, it does not account for $\epsilon_N$, and it does not account for the stochastic effects of the ALPs. In addition, Gramolin *et al*. (*61*) have recently independently shown how to account for the stochastic properties of the ALPs, in a method that is similar to our detailed analysis in (*41*).

## MATERIALS AND METHODS
### The Floquet bandwidth enhancement

The Floquet field $B_F \cos(2\pi f_F t)\hat{z}$ is aligned with the previously described axial field, $B_{ext}^z$. The strong field modulates the ground-state energy of the rubidium spins, thereby dressing their energy levels with the field induced by the radio frequency (RF) photons (*62*). This modulation is spectrally manifested as a series of resonance bands that appear at discrete harmonics of the driving field in the magnetic spectrum of the rubidium, and via multiphoton processes, it encodes the response of the spins to low-frequency fields. The resulting tilt of the measured rubidium is thus obtained from Eq. 5 by shifting the spectrum by $nf_F$ with an integer $n$ and multiplying the response with a bandwidth enhancement factor $\eta_F^{(n)}$ such that around the harmonics of the Floquet frequency $\theta_{Rb}^{Floquet}(f = f_{DM} + nf_F) = \eta_F^{(n)}\theta_{Rb}(f = f_{DM})$.

In Fig. 4A, we exemplify the Floquet spectrum via typical measurement of the spectrum, $\theta_{Rb}(f)$, in response to a low-frequency transverse magnetic noise, whose variance is white up to a cutoff frequency set at $f_c = 10$ kHz (light green shade), while setting $f_{Rb} = 105$ kHz. In the absence of the Floquet field, the rubidium response appears at the same frequencies of the drive (green line), and the response is suppressed because the driving field is tuned away from resonance $|f_{Rb} - f_c| \simeq 14\Gamma_{Rb}$. By applying a strong and resonant Floquet field ($f_F = f_{Rb}$, $B_F = 0.4$ G), the low-frequency spectrum is mapped to discrete harmonics of the Floquet field at integer multiples $n$ for which efficient coupling is realized. We observe an $\eta_F^{(1)} = 5.9$-fold enhancement of the magnetic response near the first Floquet band with respect to the low-frequency response of the unmodulated sensor. Notably, routine calibrations of the Floquet parameters during the search did not inject magnetic noise but used coherent sinusoidal signals, following the protocol that is described in detail in (*41*).

The resonant nature of the Floquet modulation is encoded in $\eta_F^{(n)}$, which depends on the external magnetic field and the Floquet amplitude, $B_F$. In Fig. 4B, we show the enhancement, $\eta_F \equiv \eta_F^{(1)}$, as a function of the axial magnetic field. At each measured value, we optimized for the frequency and amplitude of the Floquet field, following the theoretical analysis detailed in (*41*).

### Background and signal sensitivity

The sensor is susceptible not only to the anomalous fields but also to various sources of noise that limit its detection sensitivity. In this section, we present and characterize the noise model for the detector, whereas the protocol that routinely monitored and calibrated for variations of the model parameters during the search is described in (*41*). The dominant sources of noise are the magnetic field noise, $\delta B$, and optical polarization noise due to the shot noise of the probe beam $W$. In the presence of the former, Eqs. 4 and 5 are modified by taking $b_{DM} \rightarrow b_{DM} + \delta B$ and $b_{Xe}\theta_{Xe} \rightarrow b_{Xe}\theta_{Xe} + \delta B$. To study the response of the system to the above noise, we combine

Eqs. 4 and 5 and rewrite the Floquet-demodulated output of the optical magnetometer $S + N$, decomposed to the noise contribution

$$N(f) = \xi \frac{\delta B(f)}{1 + i(f_{Xe} - f)/\Gamma_{Xe}} + \delta B(f) + W(f) \qquad (6)$$

and the coherent signal contribution of the ALP at $f = f_{DM}$

$$S(f) = \xi \frac{b_{DM}}{1 + i(f_{Xe} - f)/\Gamma_{Xe}} \qquad (7)$$

The detector reading, $S + N$, is measured around the first Floquet band and is given here in units of magnetic field. The calibration protocol of the magnetometer, which determines the proportionality constant that converts the measured optical signal to magnetic field units, is described in section S3 in (*41*). Here, $W$ denotes the optical probe noise obtained after demodulation. $\xi = \gamma_{Xe}b_{Xe}/2\Gamma_{Xe}$ is the overall dimensionless factor that encodes the enhancement of the rubidium response to the xenon precession over the direct impact of magnetic noise. This is in line with the view of the xenon precession as the signal (also affected by regular and anomalous magnetic fields) measured via the optical magnetometer. $\xi$ is calibrated routinely, and during the entire search, its value ranged from 1 to 3 [see (*41*) for further details]. For $\xi \gg 1$ and $\xi|\delta B| \gg |W|$, the experimental setup reaches a maximal sensitivity to the ALP couplings ($g_{aNN}$) and is limited solely by the magnetic field noise and the gyromagnetic ratio of the nuclei. In the absence of a noise-cancellation mechanism, this limit is universal to all types of NMR sensors, independent of the number of polarized nuclear spins, the performance of the used magnetometer, or the coherence times of the atoms.

To exemplify the noise characteristics of the detector, in Fig. 5, we present the square root of the spectral density of a noise measurement for a single recording at $B_{ext}^z = 0.1$ G. In this measurement, no coherent calibration signals (other than the Floquet drive) are present, and the spectral density of the recorded noise realization PSD($N$) is calculated using Welch's method. The noise spectrum has a typical square root of spectral density of 100 fT/$\sqrt{\text{Hz}}$ (dark blue shaded region) and is dominated by noise of the probe beam $W$, which we show in (*41*) to be governed by photon shot noise. The estimated contribution of the magnetic noise is $\approx 10$ fT/tHz, generated by the inner layer of the magnetic shield (*63*) and sensed by the rubidium (light blue region) and xenon (blue region) spins. We find that only at a few frequencies (and, in particular, at harmonics of the mains hum) the magnetic noise becomes dominant over the probe noise.

To characterize the response to transverse fields, we inject white magnetic noise along the $\hat{x}$ direction with a spectral density of 5.6 pT/$\sqrt{\text{Hz}}$ as presented in Fig. 5B. To highlight the response of the xenon nuclei, we tune the Floquet parameters of the rubidium vector magnetometer to be most sensitive to fields along the $\hat{y}$ direction, suppressing the response to $\hat{x}$ fields. Thus, we gain sensitivity to the response of the xenon spins to the injected noise while suppressing its direct effect on the rubidium magnetometer. The Lorentzian response of the xenon near its NMR frequency validates Eq. 7 and enables also to estimate its parameters, yielding $\xi = 2.2$. The injection of magnetic noise to characterize the detector was used only for the particular demonstration in Fig. 5B to highlight the coupling of magnetic noise and its decomposition to contributions of Rb and Xe. The routine calibration protocols that calibrate $\xi$ use coherent sinusoidal signals instead as described in (*41*).

The detailed and strict search procedure and the methodology of the data analysis are described in (41), taking into account the exact measurements of the search and the statistical properties of the ALPs. Nonetheless, we find it insightful to exemplify the spectral sensitivity to anomalous fields of the sensor based on a short measurement of the sensor in the limit of $\nu_{vir} \to 0$ as shown in Fig. 5C. It is apparent that, when white noise dominates the spectrum, the sensitivity follows the Lorentzian NMR shape of the nuclear spins, and a high sensitivity is limited to a narrow frequency band, whose width is determined by the xenon's spectral width $\Gamma_{Xe} = 0.3$ Hz.

It is also possible to give an approximate estimate of the excluded $b_{DM}$ and the typical $\theta_{Xe}$ from the measurement in Fig. 5A at $f_{DM} \approx$ 116 Hz in resonance with the NMR frequency. For the characteristic noise spectral density $\sqrt{PSD(N)} \approx 100$ fT/$\sqrt{(Hz)}$, a finite measurement interval $T \approx 3000$ s, which is shorter than the ALP coherence time, and $\xi = 2.2$, we can exclude coherent anomalous fields that are larger than $b_{DM} \gtrsim \sqrt{PSD(N)}/(\xi\sqrt{T}) \approx 1$ fT. This field also corresponds to a minimal tilt of the Xe spins by about $\theta_{Xe} \gtrsim 2 \times 10^{-8}$, obtained by Eq. 4 for $\Gamma_{Xe} = 0.3$ Hz. Note that these approximate estimates compared the power of the signal to the noise variance to determine exclusion [signal-to-noise ratio (SNR) = 1], whereas actual constraints presented in Fig. 2 effectively set higher SNR to obtain the 95% C.L. in the likelihood test.

## Data processing

We automatically excluded data from our analysis if it matched at least one of the following criteria (quality cuts): (i) substantial variation in the sensor parameters between two subsequent calibration measurements, (ii) saturation of the detector that is identified via substantial decrease in the noise variance, and (iii) a substantial increase of transient noise, which is identified in spectral regions of the signal being far away from the NMR resonance. For (i), we veto the entire measurement, whereas for cuts related to both (ii) and (iii) data vetoing is limited to a finite measurement window.

We analyze the data using the log-likelihood ratio test to constrain the presence of ALPs at frequency $f_{DM}$ with a width determined by the signal coherence time and the effects of Earth's rotation on the sensitive axes of the detector. Each measurement was used to constrain a frequency range of width 2 to 6 Hz around the NMR frequency. Bounds were set on ALP-neutron interactions, as well as quadratic interactions (13) with neutrons. To accurately account for the velocity distribution of the DM (see Eq. 1), we followed the suggested analysis in (19). We used the asymptotic formulas in (64) for the distribution of the log likelihood. For each measurement, the noise was assumed to be white and was estimated using sideband analysis in the frequency domain away from the NMR resonance.

All analysis procedures and cuts were designed in a blinded fashion, and decided in advance before looking at the data, to eliminate bias. However, after unblinding, we found that less than 0.1% of the spectral domain of the search range was statistically inconsistent with the white noise model. For this part, we have refined the statistical tests to treat transient and coherent magnetic noise. All data are found consistent with the refined model. The ALP stochastic properties, the statistical analyses, and post-unblinding changes are detailed in (41).

## SUPPLEMENTARY MATERIALS

**Acknowledgments:** We thank M. Lisanti, M. Moscella, and W. Terrano for helping us understand and validate the correct treatment of the stochastic nature of ALPs. I.M.B. is grateful for the support of the Alexander Zaks Scholarship, The Buchmann Scholarship, and the Azrieli Foundation. We thank Hu *et al.* (*43*) for useful conversations regarding their work. **Funding:** T.V. is supported by the Israel Science Foundation-NSFC (grant no. 2522/17), by the Binational Science Foundation (grant no. 2016153), and by the European Research Council (ERC) under the EU Horizon 2020 Programme (ERC-CoG-2015, proposal no. 682676 LDMThExp). **Author contributions:** All authors contributed to the experimental design, construction, data collection, and analysis of this experiment. R.S. claims responsibility for Fig. 1. I.M.B. claims responsibility for Figs. 2 and 3 and fig. S4. G.R. claims responsibility for Figs. 4 and 5 and figs. S1 to S3. **Competing interests:** The authors declare that they have no competing interests. **Data and materials availability:** All data needed to evaluate the conclusions in the paper are present in the paper and/or the Supplementary Materials. Search datasets have been deposited in the Zenodo online repository (*65*) and in GitHub (*48*).

Submitted 11 August 2021
Accepted 13 December 2021
Published 4 February 2022
10.1126/sciadv.abl8919

# Science Advances

## New constraints on axion-like dark matter using a Floquet quantum detector

Itay M. BlochGil RonenRoy ShahamOri KatzTomer VolanskyOr Katz

**View the article online**
https://www.science.org/doi/10.1126/sciadv.abl8919
**Permissions**
https://www.science.org/help/reprints-and-permissions

Use of think article is subject to the Terms of service

Itay M. Bloch, Gil Ronen, Roy Shaham, Ori Katz, Tomer Volansky, Or Katz*

*Corresponding author. Email: or.katz@duke.edu

**This PDF file includes:**

# S1. DETAILED EXPERIMENTAL CONFIGURATION

We use a cubical borosilicate cell of length $l = 4$ mm containing 40 Torr of isotopically enriched $^{129}$Xe gas and a droplet of naturally abundant rubidium metal. The cell also contains 40 Torr of $N_2$ gas for quenching and mitigation of radiation trapping. We use twisted-pair resistance wires with current oscillating at 512 kHz to stabilize the cell temperature to $T = 150\,°$C, yielding an estimated rubidium density of $n_{Rb} \approx 1 \times 10^{14}$ cm$^{-3}$. We shield the cell from external magnetic fields by using five layers of concentric $\mu$-metal shields. We use three sets of coils to control the magnetic field: 4-winding double Helmholtz coils for controlling the field along the $\hat{z}$ direction and a bird-cage coil for the transverse fields. The shields are degaussed and the transverse magnetic field is zeroed.

The Floquet field was applied using a separate, double Helmholtz coil which was positioned in close proximity to the cell. The coil length was kept short to maintain a relative constant phase of the current for the experimental frequency range. For each magnetic field setting, the Floquet frequency was tuned near the resonance frequency of the $^{85}$Rb vapor.

The optical-pumping light originates from a single-mode, linearly-polarized, distributed Bragg reflector (DBR) laser at 795 nm. It is amplified with a commercial fiber-coupled tapered amplifier, stabilized to 500 mW and passes through a $\lambda/4$ retarder to render its polarization circular. The pumping light is detuned from the $D_1$ optical line by 22 GHz, (which has a broadening of $\Gamma_{D1} = 2$ GHz due to pressure broadening) and fills the entire cell (estimated beam waist diameter of 15 mm) these measures increase the spatial homogeneity of the alkali spin polarization within the cell.

We polarize the xenon spins along $\hat{z}$ using spin-exchange optical pumping (SEOP). In the pressure regime of the experiment, the polarization process has significant contributions from both the binary and three body collisions of the rubidium with xenon atoms. Additionally, we estimate that both $^{85}$Rb and $^{87}$Rb isotopes of polarized rubidium contribute to the SEOP process.

The probe beam originates from another single-mode DBR laser at 795 nm. It is linearly polarized and oriented along $\hat{x}$ it has a Gaussian shape with waist 3 mm, and 25 mW in power. The beam is detuned 10 GHz from the resonance of the optical line yet it is attenuated by about a factor of 3 due to reflection from the optically-uncoated vapor cell and mainly from absorption by the Rb vapor. We split the outgoing 8 mW probe signal using a NPBS and feed about 4 mW of probe power into two nearly identical polarization rotation sensors. By distributing the beam to two sensors, we allow higher beam intensities while still avoiding photodetector saturation. Sensor duplication also supports the robustness of the measurement over long time periods. Each of the two sensors is set up in a homodyne detection setup comprising a zero-order $\lambda/2$ retarder at 22.5°, a PBS, and a differential photodetector. The detector outputs are subtracted and amplified. The BPD output is fed into a Lock-in amplifier (LIA) whose reference signal is the Floquet modulation signal probed in parallel to the Floquet coil. The phase of the LIA is chosen to maximize the sensitivity to the field along the $\hat{y}$ direction at the first Floquet band. The bandwidth of the LIA was set to be 5kHz which suffices in order to observe the measured signal ($f_{DM} < 1$kHz).

We independently measure acoustic noise with an acoustic sensor attached near the probe and also monitor the variation of the pump power before the entrance to the cell. These measurements allow us to study time dependent background properties of our system.

We have accurately calibrated the magnetic field $B_c$ generated by each set of coils as a function of their current $I_c$. For this calibration, we optically pumped the xenon spins for a few seconds and tilted them off the $\hat{z}$ axis with a short transverse magnetic pulse. We then recorded their precession frequency $f_{Xe}$ and fitted it to the linear dependence $f_{Xe}(I_c) = f_0 + A_c I_c$, where the small offset $f_0$ depends predominantly on the residual magnetic field and alkali spin-exchange field. Comparison of this linear dependence with $f_{Xe} = \gamma_{Xe} B_c$ completes the calibration, yielding the linear dependence $B_c = (A_c/\gamma_{Xe})I_c$. This procedure was repeated for all coils used in the setup.

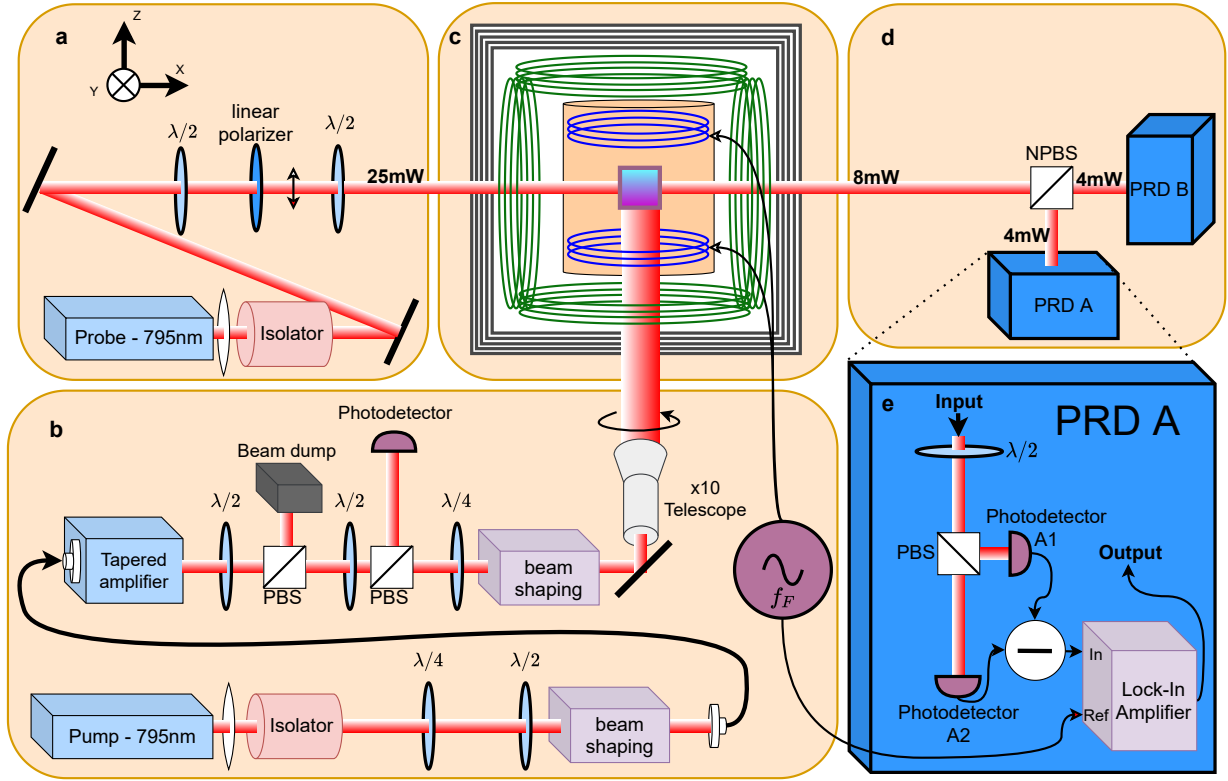

FIG. S1. **Experimental setup. a.** The probe beam originates from a 795nm free running laser diode, collimated and passes an optical isolator. a half-wave plate and a linear polarizer set the beam power and its polarization linear. **b.** The pump originates from a different free running laser diode, which is fiber-coupled to a tapered-amplifier. The beam is shaped, its polarization is set circular and time-dependent fluctuations in its power are monitored before entering the cell using a photodiode. PBS- polarization beam splitter **c.** The physical package of the detector is composed of 5 layers of $\mu$ - metal magnetic shields (black), three sets of DC coils (green) and an additional coil for the Floquet field (blue), mounted within an aluminum oxide oven (orange). **d.** The rubidium atoms partially absorb the probe beam, and the outgoing 8 mW probe intensity is splitted using a non-polarizing beam splitter (NPBS) into two polarization rotation detectors (PRD), measuring 4 mW each. The two PRD provide redundancy of the measured signal over the long measurement time, as well as intensity distribution among the two detectors to avoid sensor saturation. **e.** Configuration of a single PRD. We use the differential signal of two photodiodes to measure the polarization rotation. The differential signal is fed into a Lock-In amplifier with the reference signal being the Floquet current which is measured independently.

## S2. FLOQUET DYNAMICS

We describe the dynamics of the xenon and rubidium ensembles using the Bloch equations of the mean-spin operators $\mathbf{I}$ and $\mathbf{S}$ respectively . Under the conditions of our experiment, the spin components $S_z$ and $I_z$ maintain a constant value, which is determined by the optical pumping and spin-exchange rates. The angular tilts of the two spin ensembles $\theta_{\mathrm{Rb}} = |S_+|/S_z$ and $\theta_{\mathrm{Xe}} = |I_+|/I_z$ are determined by the transverse components of these spins $I_+ = I_x + iI_y$ and $S_+ = S_x + iS_y$ whose dynamics is described by

$$\partial_t I_+ = [i\gamma_{\mathrm{Xe}}(B_z + B_{\mathrm{F}}\sin(2\pi f_{\mathrm{F}}t)) - 2\pi\Gamma_{\mathrm{Xe}}]\, I_+ + i\gamma_{\mathrm{Xe}}I_z \tilde{B}_+^{\mathrm{Xe}}, \tag{S1}$$

$$\partial_t S_+ = [i\gamma_{\mathrm{Rb}}(\tilde{B}_z + B_{\mathrm{F}}\sin(2\pi f_{\mathrm{F}}t)) - 2\pi\Gamma_{\mathrm{Rb}}]S_+ + i\gamma_{\mathrm{Rb}}S_z \tilde{B}_+^{\mathrm{Rb}}, \tag{S2}$$

$B_z$ is the effective longitudinal magnetic field experienced by the xenon, which takes into account the applied magnetic field and the small spin exchange field exerted by the polarized rubidium spins. The effective magnetic field sensed by the alkali is denoted by $\tilde{B}_z$ and consists of the applied magnetic field, and the spin exchange field of the xenon. The inhomogeneous terms

$$\tilde{B}_+^{\mathrm{Xe}} = b_{\mathrm{DM},x} + \delta B_x + B_x^{\mathrm{cal}} + ib_{\mathrm{DM},y} + i\delta B_y + iB_y^{\mathrm{cal}} + b_{\mathrm{Rb}}\frac{S_+}{S_z}, \tag{S3}$$

$$\tilde{B}_+^{\mathrm{Rb}} = B_x^{\mathrm{cal}} + \delta B_x + i\delta B_y + iB_y^{\mathrm{cal}} + b_{\mathrm{Xe}}\frac{I_+}{I_z}, \tag{S4}$$

consist of the transverse components of the anomalous and regular magnetic fields as well as the spin exchange fields $b_{\mathrm{Rb}}S_+/S_z$ and $b_{\mathrm{Xe}}I_+/I_z$, which are the effective transverse magnetic fields the two elements induce on each other due to spin-exchange collisions. The fields $B_x^{\mathrm{cal}}$ and $B_y^{\mathrm{cal}}$ are magnetic fields which we apply through our coils only during calibration sequences.

In the experiment, we consider the spectral response of the noble gas spins near their resonance frequencies $f \approx f_{\mathrm{Xe}}$. Around that frequency, we can adiabatically substitute the solution for $S_+$ into Eq. (S1). This gives rise to corrections for $f_{\mathrm{Xe}}$ and $\Gamma_{\mathrm{Xe}}$. For our experimental conditions, both corrections can be bounded by a small correction up to $b_{\mathrm{Rb}}b_{\mathrm{Xe}}\gamma_{\mathrm{Rb}}\gamma_{\mathrm{Xe}}/2\pi\Gamma_{\mathrm{Rb}} \approx 10^{-2}$ Hz. Therefore, in the analysis we can substitute $b_{\mathrm{Rb}} = 0$ and account for the dominant effect of the alkali spins on the noble gas by using the measured $f_{\mathrm{Xe}}$ and $\Gamma_{\mathrm{Xe}}$. Additionally, the rapidly oscillating Floquet field is far off resonance from the nuclear spin, therefore has negligible direct effect on the noble-gas dynamics. In this regime, the noble gas spins near the NMR frequency is given by

$$I_+(f) = \gamma_{\mathrm{Xe}}I_z \tilde{B}_+^{\mathrm{Xe}}(f)/[(f - f_{\mathrm{Xe}}) - i\Gamma_{\mathrm{Xe}}]. \tag{S5}$$

The alkali spins in contrast, are efficiently tilted by the noble-gas spins via a combination of the fast Floquet field term and the effective transverse field $\tilde{B}_+^{\mathrm{Rb}}$, which precesses slowly with respect to the alkali resonance frequency. In this regime, substituting Eq. (S5) in Eq. (S4) and setting $b_{\mathrm{Rb}} = 0$, yields the simplified expression for the total effective transverse field measured by the rubidium

$$\tilde{B}_+^{\mathrm{Rb}} = (B_x^{\mathrm{cal}} + \delta B_x + i\delta B_y + iB_y^{\mathrm{cal}})\left(1 + \frac{b_{\mathrm{Xe}}\gamma_{\mathrm{Xe}}}{(f - f_{\mathrm{Xe}}) - i\Gamma_{\mathrm{Xe}}}\right) + \frac{b_{\mathrm{Xe}}\gamma_{\mathrm{Xe}}}{(f - f_{\mathrm{Xe}}) - i\Gamma_{\mathrm{Xe}}}(b_{\mathrm{DM},x} + ib_{\mathrm{DM},y}). \tag{S6}$$

To solve the equation of motion of the alkali spins and derive their response, we move to a frame which is rotating with the rubidium spins with an (instantaneous) complex rate

$$\tilde{\omega}(t) = 2\pi f_{\mathrm{Rb}} + \gamma_{\mathrm{Rb}}B_{\mathrm{F}}\sin(\omega_{\mathrm{F}}t) + 2i\pi\Gamma_{\mathrm{Rb}}. \tag{S7}$$

We define the spin component in that frame by $\tilde{S}_+(t) = e^{-i\int \tilde{\omega}(t)dt}S_+$. For this rotated spin, the equation of motion is

$$\partial_t \tilde{S}_+(t) = \gamma_S S_z e^{-i\int \tilde{\omega}(t)dt}\tilde{B}_+^{\mathrm{Rb}}. \tag{S8}$$

Using the Jacobi–Anger expansion

$$e^{\pm i\beta\cos(\omega t)} = \sum_{n=-\infty}^{\infty} (\pm i)^n e^{in\omega t}J_n(\beta), \tag{S9}$$

we find the solution for $\tilde{S}_+(t)$ in Eq. (S8) which is given by

$$\tilde{S}_+(t) = i\gamma_{\mathrm{Rb}}S_z e^{-i2\pi f_{\mathrm{Rb}}t + \Gamma t}\left[\sum_{n=-\infty}^{\infty}(i)^n e^{i2\pi n f_{\mathrm{F}}t}\frac{J_n(\beta)}{i(nf_{\mathrm{F}} - f_{\mathrm{Rb}}) + \Gamma_{\mathrm{Rb}}}\right]\tilde{B}_+^{\mathrm{Rb}}, \tag{S10}$$

where $\beta = \gamma_{\mathrm{Rb}} B_{\mathrm{F}}/\omega_{\mathrm{F}}$ is the modulation index of the magnetic field. Transforming back to the lab frame yields the solution with an infinite series of harmonics of the Floquet field:

$$S_+ = i\gamma_{\mathrm{Rb}} S_z \tilde{B}_+^{\mathrm{Rb}} \sum_{k=-\infty}^{\infty} c_k e^{i2\pi k f_F t}, \tag{S11}$$

with coefficients:

$$c_k = \sum_{n=-\infty}^{\infty} \frac{i^{2n-k} J_n\left(\beta\right) J_{k-n}\left(\beta\right)}{i\left(n f_{\mathrm{F}} - f_{\mathrm{Rb}}\right) + \Gamma_{\mathrm{Rb}}}. \tag{S12}$$

With the aid of the Floquet field, an infinite series of bands are formed which now contain the information of $I_+(f)$ which is usually encoded at low frequencies. Crucially, the Floquet field itself does not excite the transverse alkali spin, as absent any transverse fields the mean transverse alkali spin vanishes. It is therefore only responsible for changing the response of the alkali spins to external fields.

In our sensor, we choose $f_{\mathrm{F}} = f_{\mathrm{Rb}}$ and measure the first side-bands with $k = \pm 1$ by demodulating the measured signal with this first harmonic of the Floquet field, while filtering all other harmonics. The major contribution of the coefficient $c_k$ comes from $n = 1$, which yields an approximate simplified expression after demodulation

$$\theta_{\mathrm{Rb}}(f + f_F) = \frac{\langle S_+ \cos\left(2\pi f_{\mathrm{F}} t + \alpha\right)\rangle}{S_z} = -\frac{\gamma_{\mathrm{Rb}}}{2\Gamma_{\mathrm{Rb}}} J_1\left(\beta\right) \left(J_0\left(\beta\right) e^{-i\alpha} - J_2\left(\beta\right) e^{i\alpha}\right) \tilde{B}_+^{\mathrm{Rb}}, \tag{S13}$$

where $\alpha$ is the relative phase between the balanced signal of the photo-detectors and the reference harmonic of the lock-in amplifier. As our probe beam measures $S_x = \mathrm{Re}\left(S_+\right)$ and we wish to maximize the overall transverse sensitivity to the $\tilde{F}_{\mathrm{S}}$ term, it would generally mean to maximize $\left|J_0\left(\beta\right) e^{-i\alpha} - J_2\left(\beta\right) e^{i\alpha}\right|$. In some cases we would choose $\alpha = \frac{\pi}{2}$ and $\beta = 1.841$ such that $J_0 - J_2$ zeros [69], which allows us to ensure the stability of our calibrations by maintaining the sensitivity along $\hat{y}$ maximal while the sensitivity along $\hat{x}$ is zeroed. For this choice of parameters, the maximal realized numerical factor is given by

$$\eta_0 = J_1\left(\beta\right) \left(J_0\left(\beta\right) + J_2\left(\beta\right)\right) = 0.368 \tag{S14}$$

And the maximally attained Floquet bandwidth-enhancement factor is

$$\eta_{\mathrm{F}} = \eta_0 \sqrt{1 + (f_{\mathrm{F}}/\Gamma_{\mathrm{Rb}})^2} \tag{S15}$$

## S3. SENSOR CALIBRATION PROTOCOL

In this section, we describe the routine calibration protocols which were used during the search.

The recorded signal of the detector at the output of the lock-in amplifier, after demodulation, is denoted by $V(t)$ and is measured with an oscilloscope. This signal is proportional to the angle $\theta_{\rm Rb}(f + f_F)$ given in Eq. (S13) plus a contribution of non-magnetic noise $\tilde{W}$ (predominantly optical probe noise). To convert the measured signal $V$ to a magnetometer reading $S + N$ with units of magnetic field [c.f. Eqs.(6-7) in Methods], we calibrate its voltage response to known magnetic fields. The signal is linearly dependent on the magnetic-fields, as given by the relation

$$V = A_x \tilde{B}_x^{\rm Rb} + A_y \tilde{B}_y^{\rm Rb} + \tilde{W},\tag{S16}$$

where the coefficients $A_x$, $A_y$ are the proportionality factors in units of [V/G] which relate the measured signal $V(t)$ with the transverse magnetic field components $\tilde{B}_x^{\rm Rb} = {\rm Re}(\tilde{B}_+^{\rm Rb})$ and $\tilde{B}_y^{\rm Rb} = {\rm Im}(\tilde{B}_+^{\rm Rb})$ where $\tilde{B}_+^{\rm Rb}(t)$ is given in Eq. (S6).

Complete characterization of the sensor therefore pertains to measurement of $A_x$, $A_y$, $f_{\rm Xe}$, $\Gamma_{\rm Xe}$, and the unitless parameter

$$\xi = \frac{b_{\rm Xe}\gamma_{\rm Xe}}{2\Gamma_{Xe}},\tag{S17}$$

which denotes the enhancement factor of the detector to the NMR response of the xenon over the magnetic response of the rubidium. $\xi$ is defined with respect to linearly polarized signals of the form $B_{\rm Xe,+}(t) = B_{\rm Xe,+,0}\sin(2\pi f t)$ in Eq. (S3), which can simulate the behavior of the ALP anomalous field.

The calibration sequence we use is shown in Fig. S2. We first determine the axial field to set the NMR resonance to its predetermined value within the search range. At the first calibration step, we tune the Floquet field frequency $f_F$, its amplitude $B_F$ and the LIA phase $\alpha$. We do so by simultaneously applying two low frequency magnetic fields $B_x^{\rm cal}\cos(2\pi f_{cx} t)\hat{x}$ and $B_y^{\rm cal}\cos(2\pi f_{cy} t)\hat{y}$ with $B_x^{\rm cal} = 40\mu{\rm G}$ and $B_y^{\rm cal} = 8\mu{\rm G}$ whose frequencies scale linearly with the NMR frequency but are always detuned from it by $|f_{\rm c} - f_{\rm Xe}| \geq 30\Gamma_{\rm Xe}$ and always satisfy $f_{\rm c} \ll \Gamma_{\rm Rb}$ for both $f_{\rm c} = f_{cx}$ and $f_{\rm c} = f_{cy}$. The relatively large amplitudes of the calibration fields (with respect to the noise) and their detuning from the NMR resonance, renders $\tilde{B}_+^{\rm Rb}$ in Eq. (S6) to depend predominantly on the calibration fields $B_x^{\rm cal}$ and $B_y^{\rm cal}$. Consequently, for this configuration, the measured signal contains two sinusoidal terms oscillating at $f_{cx}$ and $f_{cy}$ with some phase shifts $\varphi_x, \varphi_y$ and is given by

$$V(t) = V_x \cos(2\pi f_{cx} t + \varphi_x) + V_y \cos(2\pi f_{cy} t + \varphi_y),\tag{S18}$$

where we can extract the measured $V_x$ and $V_y$ simultaneously, and find the coefficients $A_x = V_x/B_x^{\rm cal}$ and $A_y = V_y/B_y^{\rm cal}$. For the measured signal, $\tilde{B}_x^{\rm Rb}$ and $\tilde{B}_y^{\rm Rb}$ contribute in quadrature, and consequently we can relate the magnitude of $\tilde{B}_+^{\rm Rb}$ in the frequency domain to the measured signal in Eq. (7) absent calibration fields through

$$|V(f)| = \sqrt{A_x^2 + A_y^2}|S(f)|,\tag{S19}$$

such that $\sqrt{A_x^2 + A_y^2}$ manifests as the proportionality constant calibrating the magnetometer response.

Using this procedure we automatically tune the Floquet parameters and LIA phase using a gradient descent technique. The figure of merit which we try to minimize for the Floquet parameters depends on the frequency regime we are working at. At $f_{\rm Xe} < 125$Hz we maximized $\sqrt{A_x^2 + A_y^2}$ to achieve maximal sensitivity, whereas at $f_{\rm Xe} \geq 125$Hz we just minimized $A_x$, which was found as a robust calibration point. The sensor response by the end of this process provides the calibration values of $A_x$ and $A_y$.

We measure the NMR parameters $f_{\rm Xe}$ and $\Gamma_{\rm Xe}$ by applying a $B_x$ pulse which tilts the xenon spins off the $\hat{z}$ axis. We then monitor the precession of the xenon spins (free induction decay) which is detected by the optical magnetometer as shown in Fig. S3a. We fit this signal to a decaying cosine function and extract the resonance frequency $f_{\rm Xe}$ and the decoherence rate $\Gamma_{\rm Xe}$.

To estimate the NMR enhancement factor $\xi$, we apply again oscillatory calibration fields with amplitudes $B_x^{\rm cal}$ and $B_y^{\rm cal}$. Now scanning their frequencies $f_{cx}, f_{cy}$ near the NMR resonance. We find a Lorentzian response as shown in Fig.S3b and as expected from Eq. (S6). To shorten the calibration time during the actual search we applied the two calibration fields only on resonance at the NMR frequency, subsequently one at a time. We then recorded the two signals

$$A_x^{\rm res} = \sqrt{(A_x + A_y \xi)^2 + A_x^2 \xi^2},$$

and

$$A_y^{\rm res} = \sqrt{(A_y - A_x \xi)^2 + A_y^2 \xi^2}.$$

These measurements enable the estimation of $\xi$ by

$$\xi = \frac{1}{\sqrt{2}} \sqrt{\frac{A_x^{\mathrm{res}\,2} + A_y^{\mathrm{res}\,2}}{A_x^2 + A_y^2} - 1}. \tag{S20}$$

We find that $\xi$ ranges from 1.0 in the low frequencies ($f_{\mathrm{DM}} \leq 20 \mathrm{Hz}$) to 3.0 in the high frequencies ($f_{\mathrm{DM}} \geq 200 \mathrm{Hz}$).

## S4. OPTICAL NOISE CHARACTERIZATION

In this section, we exemplify the conversion between our measured noise to its effective magnetic noise $N$, demonstrate that our noise spectrum is dominated by noise associated with the optical probe, and that specifically, it is photon shot noise.

Using the proportionality coefficients $A_x, A_y$ we can use the proportionality constant in Eq. (S19) and convert the units of the measured noise process [which is $V(f)$ *absent calibration signals*] to its equivalent magnetic field noise whose magnitude is given by the relation

$$|N(f)| = \frac{|V(f)|}{\sqrt{A_x^2 + A_y^2}}. \tag{S21}$$

For example, for the measurement taken at Fig. 5 in the Methods section, we measure a noise spectrum of $\sqrt{\mathrm{PSD}\,(\mathrm{V})} = 8 \times 10^{-4} \mathrm{V}/\sqrt{\mathrm{Hz}}$ for a 4 mW probe beam (c.f. Fig. S1) at $f = 115 \mathrm{Hz}$, absent any coherent calibration signals. Using the procedure in section S3, we measured the coefficients $A_x = 1.2 \times 10^5 \mathrm{V/G}$ and $A_y = 7.7 \times 10^5 \mathrm{V/G}$. Substitution of these numbers in Eq. (S21) yields $\sqrt{\mathrm{PSD}\,(\mathrm{N})} \approx 100 \mathrm{fT}/\sqrt{\mathrm{Hz}}$.

We now discuss further the decomposition of the noise for its different contributions. We first characterize the system by measuring the noise of the system in the presence of the probe beam in two configurations, one in which the pump is turned off and another in which the pump is on. When the pump is off, the detector is susceptible to noise of the probe beam and electronic noise (from the instrumentation), but is insensitive to neither magnetic fields nor to the xenon precession, since the alkali spins are unpolarized. We have independently verified that pumping by the probe beam due to its residual circular polarization is more than three orders of magnitude weaker than optical-pumping by the pump. When the pump is turned on, the system becomes sensitive again to both xenon precession and magnetic fields. We found that the noise reading of the two configurations in most parts of the spectrum is similar, suggesting that the noise in these parts is dominated electronic or optical noise of the beam, and not due to processes associated with the response of spins to fields [i.e. that $W$ is the dominant term in $N$ in Eq. (7) in the Methods section]. The noise difference between the two configurations is manifested only at small parts of the spectrum, where the noise has few semi-coherent peaks (e.g. near 64 Hz which is the refresh rate of our laboratory monitors). We also note that absent the probe beam, the electronic noise is negligible.

To characterize our probe noise further, before the search we record the detector noise directly at the output of the differential photo diode and denote it by $\bar{V}(t)$, when the pump is turned off, as a function of the optical probe power at the detector $P_{\mathrm{probe}}$. Note that $V(t)$ in Eq. (S19) is demodulated and amplified by the lockin amplifier while $\bar{V}(t)$ is not, and therefore its voltage calibration is different; We convert its spectral density instead to $\mathrm{W}/\sqrt{\mathrm{Hz}}$ by the conversion coefficient $G_W = 2.75 \times 10^5 \mathrm{V/W}$ of our differential photo-detectors (after an internal trans-impedance amplifier). In Fig. S4, we present the spectral density of the measured probe noise $\mathrm{PSD}(\bar{V}/G_W)$ as a function of the probe power at an indicative frequency 160 KHz (note that the relevant detection spectrum corresponds to frequencies up to about 0.4 MHz, since $\bar{V}$ corresponds to the detector signal before the demodulation in the LIA by the Floquet frequency). We compare this measurement with the calculated Photon Shot Noise spectral density, denoted by PSN, as a function of the probe power. For our differential measurement configuration it is given by [46]

$$\mathrm{PSN} = 2h\frac{c}{\lambda}P_{\mathrm{probe}} \tag{S22}$$

where $\lambda = 795 \mathrm{nm}$ is the probe wavelength, $c$ is the speed of light and $P_{\mathrm{probe}}$ is in Watt. We find good agreement between the measurements and the photon-shot noise contribution, demonstrating therefore that it is indeed the dominant noise mechanism of our detector.

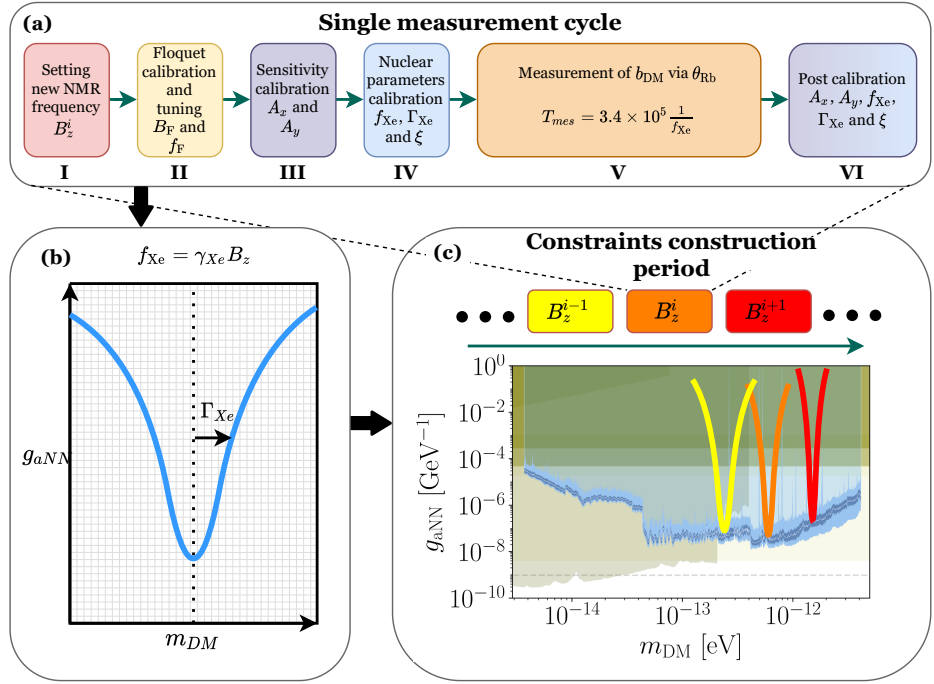

FIG. S2. **Search protocol and calibration sequence a.** Shows the series of calibration and measurement steps taken to obtain a single result. **I)** At first, The axial magnetic field is set to determine the NMR frequency and thus the central frequency of our measurement. **II)** Then the Floquet magnetometer parameters are tuned in order to achieve maximal transverse sensitivity ($f < 125$Hz) or minimal sensitivity in the $\hat{x}$ direction ($f > 125$Hz). Then follows a calibration of the alkali magnetometer **III)** and of the nuclear parameters **IV)**, as seen in Fig. S3. **V)** After calibration is done, we perform a recording of the alkali signal which spans over a substantial fraction of the ALPs coherence time. **VI)** We then perform another calibration of the magnetic and nuclear properties (leaving the Floquet parameters and magnetic fields unchanged) in order to verify the consistency of our measurements. **b.** Shows the exclusion plot derived from a single measurement. The lowest bound is achieved in near the NMR frequency $f_{Xe}$. The HWHM of the exclusion profile is $\Gamma_{Xe}$. **c.** Shows how over the course of 5 months we have assembled approximately 3000 different measurements into a single exclusion profile such as in Fig. 2. Measurements were taken to be at most $\sim 2\Gamma_{Xe}$ apart. That way each frequency has at least $\sim 1$ measurement with a resonance which is less than a $\Gamma_{Xe}$ away from the central frequency.

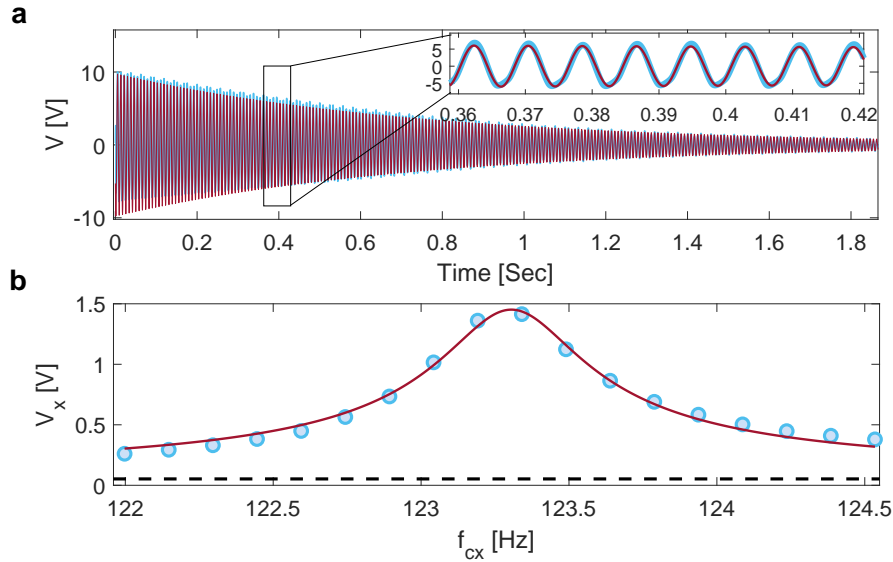

FIG. S3. **Estimation of the xenon NMR parameters. a.** Shows the response of xenon to a short transverse pulse as a function of time for an axial field of $B = 104.4$mG. From this signal we can extract the oscillations frequency $f_{Xe}$ and the decoherence rate $2\pi\Gamma_{Xe}$. **b.** Shows the nuclear response to $B_x$ as a function driving frequency. Since the optical magnetometer is tuned to be most sensitive along $B_y$ the measurement highlights the response of the Xe spins over the direct excitation of the rubidium. We can verify that the response profile is indeed a lorentzian and use it to extract $\xi$. The residual alkali response to magnetic excitations in the x axis causes the lorentzian to have a slightly elevated offset.

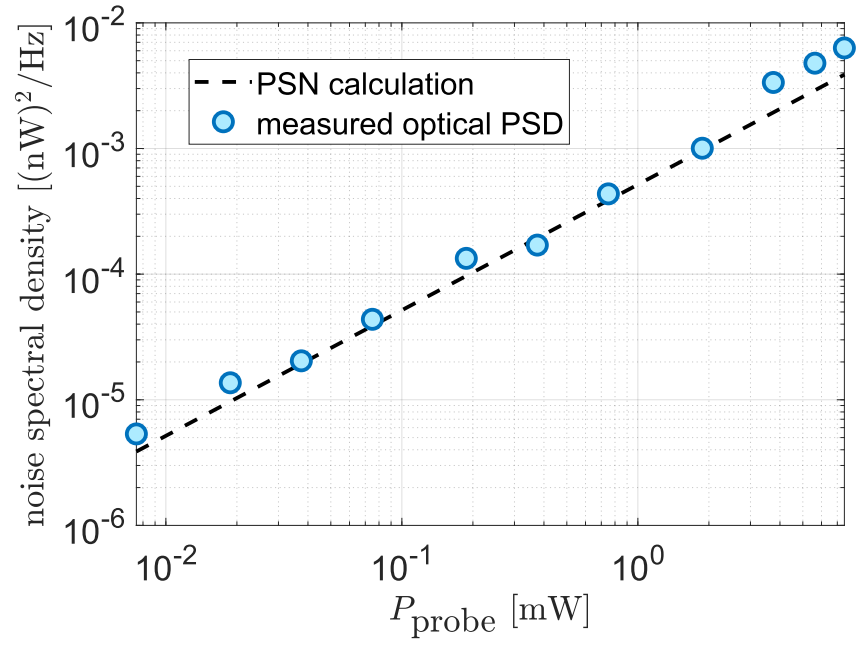

FIG. S4. **Identification of the optical probe noise with photon shot noise**. The spectral density of the probe beam absent signal (circles) at an exemplary frequency $f = 160$ KHz, compared with theoretical calculation of photon shot noise (PSN) for this experimental configuration [c.f. e.g. S22] (dashed line). The measurement is read directly after the balanced differential photodetector, before demodulation filtering and further amplification by the lock-in amplifier, and in the absence of the pump beam.

## S5. DATA AND EXCLUSION CRITERIA

### A. The Data

The optical signal imprinted on the polarization of the probe beam is measured using a Homodyne detection setup. For each photodiode, signals were recorded with a set of two oscilloscopes, whose data acquisition settings are tuned such that at any given time at least one of the oscilloscopes is recording the data. We synchronize the temporal readings of the two oscilloscopes using a pulsed sinusoidal reference signal. Since the data is filtered from high frequencies, we are able to use Shannon-Whitticker sinc interpolation to synchronize the two scopes to an accuracy of better than $10\ \mu s$. Measurements at different magnetic field configurations were recorded with timestamps accurate to about 30 seconds.

All measurements were performed at a laboratory located at $33°$N $35°$E, and the pump beam was pointing parallel to the ground along $\hat{z}$, which pointed $325°$ with respect to east (i.e. at North-North-West). The detector is mostly sensitive to anomalous fields in the transverse plane to the pump ($xy$).

As the data was collected in a period of several months, some changes were introduced in the measurement hardware and procedures as follows. The measurements during the first month used the primary homodyne detection setup as the sole photodiode. Second homodyne detection was added by the end of the first month. This gives a second independent measurement of the Photon Shot Noise, and is also used as a backup measurement setup. The absolute majority of measurements measured for roughly $3.4 \times 10^5$ oscillations of the resonance frequency. Five long measurements in the search range of $10-20$ Hz recorded $1.8 \times 10^6$ oscillations of the central frequency of search. In these measurements (and an additional one of standard length), the automatic calibration protocols (taking roughly 7 minutes) were performed following every 60 minutes of recording to ensure long-term stability. Additionally, 17 measurements in the search range of $3-11$ Hz were taken in the absence of the Floquet field and measured the unmodulated response near $n=0$ directly. Each of these experiments recorded data for four hours corresponding to about $(1-10) \times 10^4$ oscillations.

The analyzed data was digitally compressed for computational efficiency by a band-pass filter using in a window of about $\pm 10$ Hz or less around the NMR frequency. The data from the two photodiodes is then blinded in a small frequency window of $2-6$ Hz around the NMR frequency whereas the rest of the window is unblinded and used to learn about the background. This frequency-filtered blinding enabled the characterization of the noise while remaining blind to the data that was used in the final analysis.

### B. Quality Cuts

To utilize the recorded data for search of Axion-like Dark Matter, it is crucial to exclude particular intervals in which the detection sensitivity has been interfered or drifted due to technical issues. To eliminate potential bias of the results, we automatically apply a set of predefined rules in the data interpretation which identify such intervals and veto them (quality cuts). These quality cuts are classified into veto of measurements whose sensor recording was impaired and cuts in which the calibration has been uncertain or drifted. We begin with the cuts related to impairing of the recorded signal or infiltration of transient noise.

- The first cut vetoes data where the recorded signal was saturated. When a photodiode, or oscilloscope is saturated, the sensitivity to both signal and noise sharply declines. We identified these events efficiently by measuring a sharp decrease in the measurement variance which was estimated in windows of 100 milliseconds. When a 100 milliseconds interval exhibits an extremely low noise (of the order of the electronic resolution of the oscilloscope) the interval is vetoed, as well as the $\sim$25 seconds before and after the interval. If possible, we used the data of the other diode-set.

- We vetoed data intervals in which an abrupt noise was measured. The main two sources of technical noise were acoustic noise and sudden small fluctuations of the optical wavelength of the pump before the tapered amplifier. The acoustic noise originated primarily from passage of trains (the detector was positioned about 100 meters from a track), vibrations of a nearby water chiller, and sporadic vibrations of the optical table by humans. To identify these events yet maintain the data blinded, we monitored the noise variance of the data notch-filtered around the frequencies under search. This cut works by dividing the data to time-windows of size $dt$, and computing the norm of the data in each time interval. We call the vector of norms $\sigma(dt)$. We normalize each norm according to the number of active data points within the time-interval. Our goal is to find specific time intervals of duration $dt$ where the norm was abnormally large compared to the rest of the norms for time intervals of the same duration. To define what is considered an abnormally large norm, we compare the components of $\sigma(dt)$ with the following threshold

$$\sigma_{\max}(N) = N \cdot (Q(\sigma(dt), 0.75) - Q(\sigma(dt), 0.5)) + Q(\sigma(dt), 0.5), \tag{S23}$$

where $Q(\sigma(dt), P)$ is the $P$th quantile of the norms, i.e. $Q(\sigma(dt), 0.5)$ is the median norm deviation and $Q(\sigma(dt), 0.75)$ is the 75% quantile of the norms. $N$ is a parameter, which controls the fraction of norms that would be considered abnormal. Larger $N$ would mean that a smaller fraction of the vector components would be considered abnormally large.

To avoid the possibility of a short but large transient noise making a longer time-interval appear abnormally large, as a first step we veto vector components of a small $dt$ (depends on the exact details and roughly $dt \approx$ ms), above $\sigma_{\max}(N = 30)$. Unlike the next utilizations of this cut, in the first step, time intervals are vetoed also for the purposes of computing later $\sigma(dt)$s.

Since different noise sources are associated with different time-scales, we found that after removing the highly localized noise, it is best to apply the cut with different $dt$s, which correspond to different time scales of noises to work best. Different $N$s are also necessary to avoid removing too much data, while still removing noisy regions. We used $dt = 0.1, 2, 5, 12, 32, 54$ seconds, and $N = 10, 8, 8, 8, 6, 5.5, 5.5$ respectively in the analysis. As seen from the control data, for roughly $0.5\%$ of the data different $dt$s and $N$s were chosen for a better exclusion of the noise. If over $90\%$ of a time-interval is already vetoed by previous cuts, we remove it all together as well.

The parameters of the cut were optimized to avoid losing significant regions of time, while significantly reducing the noise in the notch-filtered data. All parameters were optimized on data that was filtered from the frequencies which are sensitive to the Dark Matter. Therefore, the optimization of the parameters was done in a blinded manner and the existence or non-existence of signal in the data did not bias the cuts.

The second type of cuts was related to avoiding calibration-uncertainties. The cuts were applied as follows:

- Exclusion of the few measurements which lacked one of the calibration sequences (either preceding or following the recording interval). In this case the entire measurement at that specific axial field was excluded. Typical measurements start and end with calibration sequences as shown in Fig. S3. We associate this problem with either power outages or communication errors.

- Exclusion of measurements in which a considerable variation between the initial (denoted with $i$) and final (denoted with $f$) calibration measurements was noted. These include variation of the linewidth by $|\Gamma_{\mathrm{Xe}}^{(i)} - \Gamma_{\mathrm{Xe}}^{(f)}| > 0.05$ Hz, variation of the NMR frequency by $|f_{\mathrm{Xe}}^{(i)} - f_{\mathrm{Xe}}^{(f)}| > 0.3(\Gamma_{\mathrm{Xe}}^{(i)} + \Gamma_{\mathrm{Xe}}^{(f)})$ and variation of the enhancement factor by $|\xi^{(i)} - \xi^{(f)}| > 0.15(\xi^{(i)} + \xi^{(f)})$. The fitted parameters are estimated from two different sets of photodiodes (used for redundancy), and therefore we had two estimations of the fit parameters for each calibration measurement. If both diodes pass the quality cut (as in most cases), we use the primary diode's calibration. If only one of the two diodes passes the quality cut, we use that specific diode's for estimation of parameters. We note that the two sets of diodes are independently calibrated, and that the measured $\xi$ are always in good agreement.

- Exclusion of a particular diode measurement if the calibrated vector-magnetometer response (Gauss to volt) varied by more than $30\%$ in amplitude or by more than $0.4$ Radians in the $xy$ plane between the initial and final calibration sequences. The calibrations of the two diode sets are independent. It is apparent mostly in the secondary set of diodes which had significantly smaller active area (and therefore higher sensitivity to angular drifts of the beam) and used a home-made demodulation hardware that is more susceptible to temperature drifts. We note also that our calibrations recorded the sensitivity without tracking relative signs (i.e. $\phi = \arctan(|A|_x, |A|_y)$. However as the drift is small (the majority of measurements are stable up to $0.2$ radians and outliers up to $0.5$ radians within a single measurement window) there is no phase ambiguity in the difference between the initial and final phases.

A small part of the measurements lasted about a day long, and to avoid potential drifts we calibrated and configured the response of the detector every hour. Cuts were taken similarly except that $i$ and $f$ denote subsequent calibration measurements and the threshold for variation of NMR linewidth was changed to $30\%$. We also aligned the vector-magnetometer direction detection of the primary diode set every calibration to ensure coherent measurement during recording, but observed overall relatively small variations for the two diodes sets.

## S6. STATISTICAL MODEL OF DARK MATTER

### A. ALP Dark Matter

In the main text we have presented the interaction of ALPs with the detector, as interactions with individual ALPs, averaged over the stochastic properties of the ALPs (e.g. Eqs. (1,2)). In this section, we explicitly discuss the stochastic model of the ALP fields, and present the formal averaging procedure based on similar methods to Ref. [19], and private communications with the authors of that paper.

While the averaged description describes the anomalous field amplitude $\mathbf{b}_{\mathrm{DM}}$ and the frequency $f_{\mathrm{DM}}$ as different variables, in the stochastic model they have statistical correlations. To define their relations, we begin our analysis by presenting the gradient of a single ALP field. In the rest frame of the sun, a single ALP particle $a(x,t)$ with a well defined velocity $v$ has the following gradient,

$$\nabla a(t)|_{\mathrm{sp}} = \sqrt{\frac{\rho_{\mathrm{DM}}}{2N_{\mathrm{DM}}}} e^{iE_{\mathrm{DM}}t/\hbar + i\phi_{\mathrm{DM}}} \mathbf{v} + h.c., \tag{S24}$$

where $E_{\mathrm{DM}} = \hbar\omega_{\mathrm{DM}}$ is the energy of the ALP, which is simply $E_{\mathrm{DM}} = m_{\mathrm{DM}}c^2 + m_{\mathrm{DM}}v^2/2$, for $m_{\mathrm{DM}}$ the mass and $\mathbf{v}$ the velocity. The amplitude is normalized by the energy density of DM, $\rho_{\mathrm{DM}}$, and the number of ALPs that can be summed coherently $N_{\mathrm{DM}} \gg 1$. In this frame of reference and for ultralight masses, the detector's position is nearly stationary and the x-dependence of the ALP phase ($ik_{\mathrm{DM}}x$ with $k_{\mathrm{DM}}$ the ALP wavenumber) is absorbed into $\phi_{\mathrm{DM}}$ which acts as a random phase.

The actual ALP field gradient is given by summing over the contribution of all single ALP fields in the same region, whose stochastic variables $\mathbf{v}$, $E_{\mathrm{DM}}$ and $\phi_{\mathrm{DM}}$ are sampled independently. This gradient field is given by

$$\nabla a(t) = \sqrt{\frac{\rho_{\mathrm{DM}}}{2N_{\mathrm{DM}}}} e^{im_{\mathrm{DM}}c^2 t/\hbar} \sum_{j=1}^{N_{\mathrm{DM}}} e^{im_{\mathrm{DM}}v_j^2 t/(2\hbar) + i\phi_{\mathrm{DM},j}} \mathbf{v}_j + h.c., \tag{S25}$$

where $\mathbf{v}_j$ are the ALP velocities, each sampled independently from the Standard Halo Model (SHM) [42], and $\phi_{\mathrm{DM},j}$ are the $N_{\mathrm{DM}}$ phases which are sampled independently from a uniform distribution between 0 and $2\pi$. We index the different ALP fields by $j \in \{1, ..., N_{\mathrm{DM}}\}$.

Assuming a finite measurement time $T$ of that field, it is fruitful to represent the time-dependent ALP signal via a Fourier series in the frequency domain

$$(\nabla a)_n = \mathbf{W}_n (m_{\mathrm{DM}}, \mathbf{\Phi}_{\mathrm{DM}}) + \mathbf{W}_n (-m_{\mathrm{DM}}, -\mathbf{\Phi}_{\mathrm{DM}}), \tag{S26}$$

where

$$\mathbf{W}_n (m_{\mathrm{DM}}, \mathbf{\Phi}_{\mathrm{DM}}) = \frac{\sqrt{2\rho_{\mathrm{DM}}}}{2\sqrt{N_{\mathrm{DM}}}} \sum_{j=1}^{N_{\mathrm{DM}}} (-1)^n e^{i\Phi_{\mathrm{DM},j}} \mathrm{sinc}\left(\frac{(m_{\mathrm{DM}}v_j^2 + 2m_{\mathrm{DM}}c^2 - 2\hbar\omega_n)T}{4\hbar}\right) \mathbf{v}_j. \tag{S27}$$

We index the discrete frequencies of the series by $\omega_n = 2\pi n/T$, and further define the vector $\mathbf{\Phi}_{\mathrm{DM}}$ whose $j$ component indexes the single ALP contribution

$$\Phi_{\mathrm{DM},j} = \phi_{\mathrm{DM},j} + E_{\mathrm{DM},j}T/(2\hbar). \tag{S28}$$

We simplify Eq. (S27) by accounting for properties of the SHM. The SHM is spherically symmetric around the sun's velocity $\mathbf{v}_\odot (= (11, 232, 7) \mathrm{km/s}$ in the galactic frame), and therefore it is useful to write

$$\mathbf{v} = v_1 \hat{v}_1 + v_2 \hat{v}_2 + (v_3' + v_\odot)\hat{v}_3, \tag{S29}$$

where $\hat{v}_1, \hat{v}_2$ are the two perpendicular directions to $\hat{v}_3 \equiv \hat{v}_\odot$. $v_1, v_2, v_3'$, are identically distributed random variables, sampled from a Gaussian distribution with a cutoff,

$$f_{\mathrm{SHM}}(\mathbf{v}) = \begin{cases} \frac{e^{-\left(\frac{\mathbf{v}-\mathbf{v}_\odot}{v_{\mathrm{vir}}}\right)^2}}{N_{\mathrm{SHM}}} & |\mathbf{v} - \mathbf{v}_\odot| < v_{\mathrm{esc}} \\ 0 & \text{else,} \end{cases} \tag{S30}$$

where the cutoff is set by the escape velocity $v_{\mathrm{esc}} = 540$ km/s. Here $v_{\mathrm{vir}} = 220$ km/s is the virial velocity, and $N_{\mathrm{SHM}}$ is a normalization factor set such that $\int d^3v f_{\mathrm{SHM}}(\mathbf{v}) = 1$. The distribution for $v_1$ and $v_2$ is symmetric around 0, and cross correlations between different axes vanish since they are always anti-symmetric in either $v_1$ or $v_2$.

To simplify the analysis, we approximate the distribution of the total ALPs field by a multi-dimensional gaussian distribution. This approximation, utilizes the Multi-Dimensional Central-Limit Theorem (MDCLT) which is valid in our ultralight mass range for which the number of ALPs composing the sum is large ($N_{\rm DM} \gg 1$). To make our equations more compact, we will use the MCDLT on the vector $\bar{\mathbf{A}}$ defined as

$$\bar{\mathbf{A}} = \nabla a / \sqrt{2\rho_{\rm DM}}, \tag{S31}$$

and index the vector components by $\bar{A}_{n,i}$ such that $n$ denotes the frequency series component and $i$ denotes the scalar field's gradient's direction using the coordinate system in Eq. (S29). Since the initial single-ALP phases are random, $\bar{\mathbf{A}}$ has a zero mean. Therefore, the distribution of $\bar{A}$ is given by

$$P_{\bar{A}}(\bar{\mathbf{A}}) = \frac{1}{(2\pi)^{n_{\bar{A}}} |\Sigma_{\bar{A}}|} e^{-\bar{\mathbf{A}}^\dagger \Sigma_{\bar{A}}^{-1} \bar{\mathbf{A}}/2}, \tag{S32}$$

where $n_{\bar{A}}$ is the number of elements in the vector $\bar{\mathbf{A}}$. $n_{\bar{A}}$ is equal to three times the number of ALP frequencies which dominate the ALPs spectrum (since there are three independent directions). We also defined the $n_{\bar{A}} \times n_{\bar{A}}$ covariance matrix of $\bar{\mathbf{A}}$ as $\Sigma_{\bar{A}}$. The determinant of the covariance matrix is denoted by $|\Sigma_{\bar{A}}|$. We now turn to compute the complex values of $\Sigma_{\bar{A}}$. Using Eqs. (S26), (S27) we find that given any $n, i$, the real and imaginary parts of any $\bar{A}_{n,i}$ are independent identically distributed variables. Since the field is real valued, the negative Fourier series components are simply the complex conjugates of the positive Fourier series components. The positive Fourier components $n, l$ of the field therefore satisfy for all $i, j$,

$$\text{Cov}(\text{Re}(\bar{A}_{n,i}), \text{Re}(\bar{A}_{l,j})) = \text{Cov}(\text{Im}(\bar{A}_{n,i}), \text{Im}(\bar{A}_{l,j})), \tag{S33}$$

and

$$\text{Cov}(\text{Re}(\bar{A}_{n,i}), \text{Im}(\bar{A}_{l,j})) = 0. \tag{S34}$$

And for positive $l, n$ and any $i, j$, the elements of the covariance matrix $\Sigma_{\bar{A}}$ are given by

$$\text{Cov}(\text{Re}(\bar{A}_{n,i}), \text{Re}(\bar{A}_{l,j})) = \frac{\delta_{ij}}{2} \int d^3v f_{\rm SHM}(\mathbf{v})(\mathbf{v})_i(\mathbf{v})_j (-1)^{n-l} \text{sinc}\left(\frac{(\omega_{\rm DM}(v) - \omega_n)T}{2}\right) \text{sinc}\left(\frac{(\omega_{\rm DM}(v) - \omega_l)T}{2}\right), \tag{S35}$$

with $\omega_{\rm DM}(v) = (m_{\rm DM}c^2 + m_{\rm DM}|\mathbf{v}|^2/2)/\hbar$. Notably, from cylindrical symmetry in the the SHM coordinates we find that $\text{Cov}(\text{Re}(\bar{A}_{n,1}), \text{Re}(\bar{A}_{l,1})) = \text{Cov}(\text{Re}(\bar{A}_{n,2}), \text{Re}(\bar{A}_{l,2}))$. Considering only positive frequencies (since the negative ones are simply the complex conjugates of the positive ones), the covariance can only be significant between two frequencies in an interval centered roughly around $f_{\rm center} \approx m_{\rm DM}(c^2 + 3v_{\rm vir}^2/2)/h$, with a width of approximately $f_{\rm width} \approx m_{\rm DM}v_{\rm vir}^2/\hbar$. Consequently, the number of frequency components needed to characterize the field distribution at a specific time interval $[0, T]$ is a few times $\max(1, m_{\rm DM}v_{\rm vir}^2 T/\hbar)$.

An example for $\text{Cov}(\text{Re}(\bar{A}_{n,i}), \text{Re}(\bar{A}_{l,j}))$ ($= (\Sigma_{\bar{A}})_{(n,i),(l,j)}$) can be seen in Fig S5. The figure shows $\text{Cov}(\text{Re}(\bar{A}_{n,i}), \text{Re}(\bar{A}_{n,i}))$, i.e. the diagonal elements of $\Sigma_{\bar{A}}$. The light blue (dark blue) points are the diagonal elements for $i = 3$ ($i = 1, 2$), which are parallel (perpendicular) to the sun's velocity in the galactic frame (c.f. Eqs. (S29,S35)). We have chosen to show the covariance elements for $m_{\rm DM}c^2 T/\hbar = 10^7$ (i.e. a measurement longer than the coherence time), such that several different frequencies carry signal. Several observations can be made by looking at the figure. First, the covariance is a few times larger for the components which are parallel to the sun. This can be easily understood since the average $|v_3|$ is larger by $\sim v_\odot$ than the average $|v_1|$ and $|v_2|$. We also note that for $\hbar \omega_n < m_{\rm DM}c^2$, the covariance is nearly 0, since no individual ALPs carry an energy that is below the rest mass. Furthermore, the covariances central frequencies are roughly $m_{\rm DM}(c^2 + 3v_{\rm vir}^2/2)/\hbar$, and there are roughly $3m_{\rm DM}v_{\rm vir}^2$ frequencies that carry a significant fraction of the ALP signal. Finally we emphasize that while the $x$-axis is shifted by $m_{\rm DM}c^2/\hbar$ for convenience, all shown frequencies are in fact positive ($\omega_n > 0$).

The ALP gradient presented here is directly related to the anomalous field $\mathbf{b}_{\rm DM}$, which appears in the averaged Hamiltonian in Eq. (1). For completeness, these two quantities are related by

$$\nabla a(t) = \frac{\gamma_{\rm Xe} \mathbf{b}_{\rm DM}}{g_{\rm a-Xe}} \cos(2\pi f_{\rm DM}t + \phi_{\rm DM})/\sqrt{N_{\rm DM}}, \tag{S36}$$

with

$$g_{\rm a-Xe} \equiv \epsilon_N g_{aNN} + \epsilon_P g_{aPP}. \tag{S37}$$

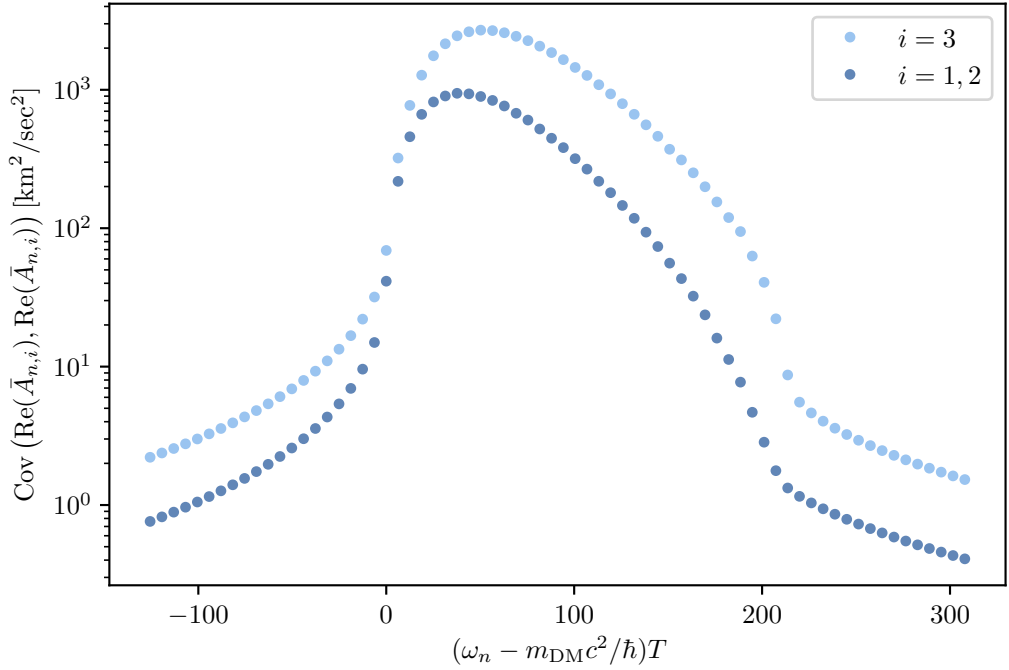

FIG. S5. **Example of the diagonal elements of the covariance matrix of $\bar{\mathbf{A}}$.** An example of the diagonal elements of the covariance matrix of the variable $\bar{\mathbf{A}}$ (which is proportional to the gradient of the ALP). The **light blue** (**dark blue**) points correspond to the elements in the diagonal of the covariance matrix of $\bar{\mathbf{A}}$, which are parallel (perpendicular) to the velocity of the sun in the galactic rest frame. The legend labels describe the index $i$ of the corresponding $\bar{\mathbf{A}}$ elements, as described in Eqs. (S29,S35). See [41] text for further details.

Here we explicitly include the contribution of both neutrons and protons to the nuclear spin, whose coupling coefficients are $g_{aNN}$ and $g_{aPP}$ are weighted by $\epsilon_N$ and $\epsilon_P$. When casting bounds on coupling of DM with neutrons, we set $\epsilon_P g_{aPP} = 0$. Similarly, when in section S8 we cast bounds on protons, we set $g_{aNN}$ to 0. While this implies the two bounds are not independent, only in the unlikely possibility that the two couplings have $\epsilon_N g_{aNN} \approx -\epsilon_P g_{aPP}$, would the two particles cause a canceling of the total effect, and our bounds would be modified in a significant manner. . We also note that for consistency, the averaging operation used in Eq. (1) must be normalized by the normalization factor $1/\sqrt{N_{\mathrm{DM}}}$ instead of $1/N_{\mathrm{DM}}$.

## B. Model of the detector signal

The detector measures the spin of the alkali in the $\hat{x}$ direction, where $\hat{x}$ is defined as the direction of the probe. However, as discussed in sections S2 and S3, the detector is sensitive to magnetic fields in the entire plane which is perpendicular to the pump beam (defined as the $xy$ plane). Using the notations of the calibration parameters defined in S3,

$$S(t) = \sqrt{2\rho_{\mathrm{DM}}} g_{a-\mathrm{Xe}} \zeta(t) \mathrm{Re} \left( \left| \frac{2\xi(A_x + iA_y)}{(i(f_{\mathrm{Xe}} - m_{\mathrm{DM}}c^2/h)/\Gamma_{\mathrm{Xe}} + 1)\gamma_{\mathrm{Rb}}} \right| \sum_{\omega_n > 0} \bar{\mathbf{A}}_n e^{i\omega_n t} (\hat{x}(t) + i\hat{y}(t)) \right), \tag{S38}$$

where $S(t)$ is the measured signal as a function of time, $\zeta(t)$ is a binary function (exclusively 0 or 1), which is 0 at times which are vetoed by the quality cuts, and 1 at the times that pass the quality cuts. $\xi$ describes the nuclear enhancement factor, and $A_x$, $A_y$ are the single-axes sensitivities of the alkali away from resonance (see section S3 for further details). The sum is only over positive frequencies, since the width of the xenon resonance is narrower than the frequency of the resonance, so only $f \approx +m_{\mathrm{DM}}c^2/h$ is enhanced (this is equivalent to taking the rotating wave approximation for the anomalous field measured when it is measured by the xenon spins). The $\hat{x}(t)$ direction is the direction of the probe beam in the galactic frame, which is time dependent due to the rotation of the earth, and is calculated using the detector's geographic coordinates and orientation [c.f. Sec. S5 A]. Similarly, the $\hat{z}(t)$ is the time-dependant direction of the pump beam in the galactic frame, and therefore trivially $\hat{y}(t) = \hat{z}(t) \times \hat{x}(t)$. Note that we've absorbed some phases in $\bar{\mathbf{A}}$ since it has a random phase. We have implicitly assumed the linewidth of the ALP to be narrow enough such that taking the system response at a frequency $f = m_{\mathrm{DM}}c^2/h$ is accurate enough despite the slightly different energies of different single ALPs.

As the measured data in the time domain is a real vector we may look exclusively at the positive frequency components in the Fourier space without losing any information (the negative frequencies' amplitudes are merely the conjugates of the positive ones). The Fourier series representation of the system response in Eq. (S38) is then given by

$$S_m = \sum_{n,i} \alpha_{mn,i} \bar{A}_{n,i},$$
(S39)

and the matrix $\alpha$ can be derived from Eq. (S38), and can be found to be

$$\alpha_{mn,i} = \left| \frac{\sqrt{2\rho_{\mathrm{DM}}} g_{\mathrm{a-Xe}} \xi (A_x + iA_y)}{T(i(f_{\mathrm{Xe}} - m_{\mathrm{DM}}c^2/(h))/\Gamma_{\mathrm{Xe}} + 1)\gamma_{\mathrm{Rb}}} \right| \int_0^T dt \zeta(t) (\hat{x}(t) + i\hat{y}(t))_i e^{i(\omega_n - \omega_m)t}.$$
(S40)

Here $i$ indexes the three directions defined in Eq. (S29). For each axis, the index $n$ denotes any of the $n_{\bar{A}}/3$ points of positive Fourier frequencies which are used to describe the ALP. As the detection configuration could introduce modulation to the ALP frequencies (e.g. by earth's rotation), additional frequencies in the data can then contain ALP information. We use the index $m$ to denote any of the $n_d$ frequencies in which the ALP could appear in the data. Notably, $n_d$ is therefore slightly larger than $n_{\bar{A}}/3$. For simplicity, from this point we treat $n, i$ as a single index such that $\alpha$ is an $n_d \times n_{\bar{A}}$ matrix.

## S7. ANALYSIS

To analyze the data of the search and constrain DM, we used a profiled likelihood ratio (PLR) test. To motivate and explain the likelihood functions we constructed, we begin in section S7 A with some important definitions. In section S7 B we discuss how the noise parameters were estimated for the likelihood computation. In section S7 C we define the likelihood of a single isolated measurement. Then, in S7 D we discuss the final likelihood used in our analysis, which utilized data from multiple measurements. After the unblinding, we made several changes to the analysis which are detailed and explained in Sec. S7 F.

### A. Definitions and procedure

To set our bounds we use the PLR method (see e.g. [64,70]). In the context of our analysis, the likelihood is a function of the observed data $\mathbf{d}$, and a set of parameters that fully specify the distributions of both the signal $\mathbf{S}$ and the noise $\mathbf{N}$. We assume the three vectors $\mathbf{d}, \mathbf{S}, \mathbf{N}$ are complex vectors of finite size.

In our experiment, given the calibration measurements discussed in S3, only a few parameters remain unknown. For ALPs, the set of parameters necessary to specify the distribution of the signal are the coupling constants and the mass. Since we assume a locally white gaussian noise, specifying the distribution of the noise requires only its variance. Given these parameters, the likelihood of a specific measured data vector $\mathbf{d}$, is the probability of observing that $\mathbf{d}$ given these parameters. Using the definition of conditional probabilities, we can therefore write a generic likelihood as

$$L(\mathbf{d}) = \int dS P_s(\mathbf{S}) \int dN P_n(\mathbf{N}) P(\mathbf{d}|\mathbf{S}, \mathbf{N}). \tag{S41}$$

Here $P_s(\mathbf{S})$ is the probability of getting a specific set of amplitudes $\mathbf{S}$, and the integration $\int dS$ is run over all possible signal amplitudes. Similarly $\int dN$ integrates over all possible noise amplitudes which are relevant to the data, and $P_n(\mathbf{N})$ is the noise probability distribution. Working in the Fourier space, since the data is the sum of the signal and the noise, the conditional probability is simply reduced to

$$P(\mathbf{d}|\mathbf{S}, \mathbf{N}) = \prod_m \delta^2(d_m - S_m - N_m). \tag{S42}$$

The product is over all Fourier components of the data, indexed by $m$, and the square of the delta function is meant to signify the usage of complex variables, i.e. $\delta^2(z) \equiv \delta(\mathrm{Re}(z))\delta(\mathrm{Im}(z))$. The signal's amplitudes have a one to one correspondence with our random variable $\bar{\mathbf{A}}$, so that

$$P_s(\mathbf{S})dS = P_{\bar{A}}(\bar{\mathbf{A}})d\bar{A}, \tag{S43}$$

where here $P_{\bar{A}}(\bar{\mathbf{A}})$ is meant to signify the distribution of all $\bar{\mathbf{A}}$s which are relevant for the measurement. Integrating over $dN$ in Eq. (S41) with substitutions of Eq. (S42) and Eq. (S43), yields

$$L(\mathbf{d}) = \int d\bar{A} P_{\bar{A}}(\bar{\mathbf{A}}) P_n(\mathbf{d} - \alpha \bar{\mathbf{A}}), \tag{S44}$$

Enabling to compute the likelihood for a given noise model with distribution $P_n$ and the measured data.

### B. Noise model

We examine the noise spectrum of the detector using control measurements which are blind to the signal. First, we measured the spectrum when the pump beam was turned off, corresponding to negligible contribution of spin signal to the detector outputs. In these measurements, we found that aside from several sharp peaks at some particular frequencies, the dominant noise source was the optical noise of the probe beam which appears white within the spectral width of ALPs. See S7 F for further discussion and caveats.

After observing these initial control measurements, it was therefore decided to use a white-noise model for the noise for each specific measurement and ALP mass examined. In theory, the computation of the variance of that distribution should be part of the likelihood profiling procedure. However, we found that under the assumption of white-gaussian noise, it is computationally more efficient (and nearly as accurate) to estimate the white noise variance by computing them from side-bands of the measurements, which were insensitive to the signal.

In practical terms, for each specific measurement, photodiode, and mass examined, we assume a distribution of the noise,

$$P_N(\mathbf{N}) = \frac{1}{(2\pi)^{n_N}|\Sigma_N|}e^{-\mathbf{N}^\dagger \Sigma_N^{-1}\mathbf{N}/2},$$ (S45)

where $n_N = n_d$ is the number of Fourier frequencies looked at for the data (since each data frequency is an independent draw of the noise). Since we assume a white gaussian noise, $\Sigma_N$ is the covariance matrix, equal to the variance times an identity matrix of size $n_N \times n_N$. The noise distribution $P_n$ we consider in the full likelihood test is determined by multiplication of the individual noise distributions which compose that test, each described by Eq. (S45).

For each measurement and mass in our search range, we have experimentally measured and estimated $\Sigma_N$ using the following protocol. We look at the variance of the $M$ frequencies directly above and below the set of $n_d$ frequencies which $\mathbf{d}$ of that measurement is composed of. We assume that the variance of these $2M$ frequencies is representative of the variance of the white noise. Importantly, the $n_d$ points which are sensitive to the signal, are blinded and not used in this estimation. We found that $M = 30$ is sufficient for our estimation, as the corresponding statistical error is small and therefore, characterizes the noise variance under the white noise assumption. In this model, we assume that the noise of the different photodiodes is uncorrelated, as the photon-shot noise is independent for the two photo-diodes.

### C. Single measurement likelihood

In this section, we derive the likelihood function of a single measurement, corresponding to a specific value of the magnetic field in the search and using the signal of a single photodiode $L_1$. Substituting Eqs. (S32,S45) in Eq. (S44) yields

$$L_1 = \int d\bar{A}\, \frac{1}{(2\pi)^{n_N+n_{\bar{A}}}|\Sigma_N||\Sigma_{\bar{A}}|} e^{-\bar{\mathbf{A}}^\dagger \Sigma_{\bar{A}}^{-1}\bar{\mathbf{A}}/2 - (\mathbf{d}-\alpha\bar{\mathbf{A}})^\dagger \Sigma_N^{-1}(\mathbf{d}-\alpha\bar{\mathbf{A}})/2}.$$ (S46)

For clarity we repeat here the dimensions of the different parameters in Eq. (S46). $\bar{\mathbf{A}}$ is a complex $n_{\bar{A}}$-dimensional vector. Correspondingly $d\bar{A}$ is a $2n_{\bar{A}}$-dimensional (real) volume element. $\Sigma_{\bar{A}}$ is an hermitian $n_{\bar{A}} \times n_{\bar{A}}$ matrix. $\Sigma_N$ is an $n_N \times n_N$ matrix (proportional to the identity matrix). $\mathbf{d}$ is a complex $n_d$-dimensional vector and always $n_d = n_N$. $\alpha$ is an $n_N \times n_{\bar{A}}$ complex matrix. We remind the reader again that $n_d \geq n_{\bar{A}}/3$.

To perform the $d\bar{A}$ integration we use the generic gaussian integral formula

$$\int d^n x\, e^{-\mathbf{x}^\dagger M \mathbf{x}/2 + \mathbf{J}^\dagger \mathbf{x}/2 + \mathbf{x}^\dagger \mathbf{J}/2} = \frac{(2\pi)^n}{|M|}e^{\mathbf{J}^\dagger M^{-1}\mathbf{J}/2},$$ (S47)

for a complex $n$-dimensional variable $x$, an hermitian $n \times n$ hermitian invertible matrix $M$, and an $n$-dimensional complex vector $\mathbf{J}$. Application of Eq. (S46) in Eq. (S47) with

$$M \equiv \Sigma_{\bar{A}}^{-1} + \alpha^\dagger \Sigma_N^{-1}\alpha,$$ (S48)

and

$$\mathbf{J} \equiv \alpha^\dagger \Sigma_N^{-1}\mathbf{d},$$ (S49)

yields the likelihood for a single measurement and photodiode

$$L_1 = \frac{1}{(2\pi)^{n_N}|\Sigma_N||\Sigma_A||M|}e^{\mathbf{J}^\dagger M^{-1}\mathbf{J}/2 - \mathbf{d}^\dagger \Sigma_N^{-1}\mathbf{d}/2}.$$ (S50)

Notably, this simple likelihood function neglects any correlations between the different measurements due to the long ALP coherence time. The full likelihood we used in the search includes such correlations as discussed in the next section.

### D. Full likelihood

Had all the measurements been independent instants of $\bar{\mathbf{A}}$, the likelihood of multiple measurements would have simply been the multiplication of the different individual likelihoods set by Eq. (S50). However, since the measurements can be correlated, the computation becomes somewhat more complicated. Generally, when two measurements, are spaced by less than the coherence time of the ALP $\sim h/(m_a v_{\mathrm{vir}}^2)$, $\bar{\mathbf{A}}$, their ALP amplitudes, which are sampled from Eq. (S32), are correlated. This is most severe when we compare the data from two photodiodes that measured simultaneously the same ALP field realization and the two $\bar{\mathbf{A}}$s are identical. In the following, we construct a likelihood test which accounts for these correlations.

### 1. Non-simultaneous measurements

Let us first address only the treatment of non-simultaneous measurements, which are spaced by less than the coherence time. We call such dependent measurements a "chain" of measurements. To construct a chain, we simply take the first chronological measurement that is sensitive to a specific $m_{DM}$, and examine if there is another measurement within a fraction $r$ of the coherence time from the ending of that measurement that is also sensitive to that mass. If there is one, we add it to the chain, and check if there is another measurement which starts less than a fraction $r$ of the coherence time from the end of the last measurement in the chain. We continue this process until there are no measurements within a fraction $r$ of the coherence time from the last measurement at the chain, and identify these measurements as a single chain. We chose $r = 2/3$ as long enough time for measurements to be independent. We treat a chain as a single large time interval, and compute the distribution of a single $\bar{\mathbf{A}}$ for that interval, $P_{chain}(\bar{\mathbf{A}})$. To compute the likelihood of a single chain, we modify the distribution of the signal (as described in the previous paragraph), as well as apply several changes in Eqs. (S40),(S39).

Instead of using a single $\alpha$ matrix as in Eq. (S39), for a chain of length $K$ (i.e. made of $K$ individual measurements), we use $K$ matrices. Therefore, the $k^{th}$ measurement in the chain would have a matrix $\alpha^{(k)}$, data vector $\mathbf{d}_k$, and noise vector $\mathbf{N}_k$ associated with it. Yet all $K$ measurements share the same $\bar{\mathbf{A}}$. Therefore, the data will be equal to

$$\mathbf{d}_k = \mathbf{N}_k + \alpha^{(k)}\bar{\mathbf{A}}, \tag{S51}$$

where the individual $\mathbf{N}_k$s are sampled from each measurement's own noise distribution given in Eq. (S45). The $\alpha^{(k)}$ matrix is given by

$$\alpha^{(k)}_{mn,i} = \left| \frac{\sqrt{2\rho_{DM}} g_{a-Xe} \xi_k((A_x)_k + i(A_y)_k)}{T_{chain}(i((f_{Xe})_k - m_{DM}c^2/(h))/(\Gamma_{Xe})_k + 1)\gamma_{Rb}} \right| \int_{t_k}^{t_k+T_k} dt \zeta_k(t)(\hat{x}(t) + i\hat{y}(t))_i e^{i(\omega_n - \omega_m)t}. \tag{S52}$$

Each matrix $\alpha^{(k)}$ describes the response of the system in the $k^{th}$ measurement of the chain to the stochastic realization of $\bar{\mathbf{A}}$ for the whole chain. Similar to $\alpha$, here we also use a reshaped version of $\alpha^{(k)}$ and consider it from now on as a matrix of size $n_{d_k} \times n_{\bar{A}}$. We denote the entire time of the concatenated measurements in the chain by $T_{chain}$. Here, the binary function $\zeta_k(t)$ is 0 for times that fail the quality cuts of the $k^{th}$ measurement and is 1 otherwise. $t_k, T_k$ are the initial and total time of the $k^{th}$ measurement respectively. All calibration variables with the $k$ subscript are those computed using the calibration measurements of the $k^{th}$ measurement in the chain.

Another necessary modification to Eqs. (S40),(S39) is related to phase uncertainty which results from our uncertainty in the construction of the chain's data. While the timing within individual measurements is determined precisely, we logged the initial time-stamps of a single measurement only to a precision of about 30 seconds. For the frequency range we consider, this uncertainty is larger than the reciprocal resonance frequency. Consequently, the imprecise timing adds a random initial phase to the ALP field at the measurement number $k$ in the chain for every $k$, which results in an effective change of $\alpha^{(k)} \to e^{i\phi_k}\alpha^{(k)}$, so that Eq. (S51) is modified to

$$\mathbf{d}_k = \mathbf{N}_k + e^{i\phi_k}\alpha^{(k)}\bar{\mathbf{A}}, \tag{S53}$$

where the $\alpha^{(k)}$s are defined in Eq. (S52) and $\phi_k$ is a uniformly distributed random phase. The final likelihood would need to be integrated over all possible values of $\phi_k$.

Therefore, for a chain of K non-simultaneous measurements, we may rewrite Eq. (S41) as

$$L_{non-sim}(\mathbf{d}_1, ...., \mathbf{d}_K) = \int d\bar{\mathbf{A}} P_{chain}(\bar{\mathbf{A}}) \prod_{k=1}^{K} \left( \int \frac{d\phi_k}{2\pi} \int dN_k P_{N,k}(\mathbf{N}_k) P_k(\mathbf{d}_k|\bar{\mathbf{A}}, \mathbf{N}_k) \right), \tag{S54}$$

We will again omit the arguments of the LHS from now on for brevity. Using the conditional probability in Eq. (S42) for each individual measurement, we can perform the $dN_k$ integrations yielding

$$L_{non-sim} = \int d\bar{\mathbf{A}} P_{chain}(\bar{\mathbf{A}}) \prod_{k=1}^{K} \left( \int \frac{d\phi_k}{2\pi} P_{N,k}(\mathbf{d}_k - e^{i\phi_k}\alpha^{(k)}\bar{\mathbf{A}}) \right). \tag{S55}$$

And if we plug in the gaussian distributions, we see that this is equivalent to

$$L_{non-sim} = \int d\bar{A} \prod_{k=1}^{K} \left( \int \frac{d\phi_k}{2\pi} \frac{e^{-(\mathbf{d}_k - e^{i\phi_k}\alpha^{(k)}\bar{\mathbf{A}})^\dagger \Sigma_{N_k}^{-1}(\mathbf{d}_k - e^{i\phi_k}\alpha^{(k)}\bar{\mathbf{A}})/2}}{(2\pi)^{n_{N_k}}|\Sigma_{N_k}|} \right) \frac{e^{-\bar{\mathbf{A}}^\dagger \Sigma_{\bar{A}}^{-1}\bar{\mathbf{A}}/2}}{(2\pi)^{n_{\bar{A}}}|\Sigma_{\bar{A}}|}. \tag{S56}$$

Performing the $d\bar{A}$ integral leads us to an equation similar to Eq. (S50),

$$L_{\text{non-sim}} = \left( \prod_{l=1}^{K} \int \frac{d\phi_k}{2\pi} \right) \frac{e^{\mathbf{J}_{\text{tot}}^{\dagger} M_{\text{tot}}^{-1} \mathbf{J}_{\text{tot}}/2 - \sum_k (\mathbf{d}_k^{\dagger} \Sigma_{N_k}^{-1} \mathbf{d}_k)/2}}{(\prod_k ((2\pi)^{n_{N_k}} |\Sigma_{N_k}|)) |\Sigma_{\bar{A}}| |M_{\text{tot}}|}, \tag{S57}$$

with

$$M_{\text{tot}} \equiv \Sigma_{\bar{A}}^{-1} + \sum_k (\alpha^{(k)\dagger} \Sigma_{N_k}^{-1} \alpha^{(k)}), \tag{S58}$$

and

$$\mathbf{J}_{\text{tot}} \equiv \sum_{k=1}^{K} e^{i\phi_k} \mathbf{J}_k. \tag{S59}$$

with

$$\mathbf{J}_k = \alpha^{(k)\dagger} \Sigma_{N_k}^{-1} \mathbf{d}_k. \tag{S60}$$

### 2. Simultaneous measurements

To account for the two simultaneous measurements of the two photodiodes, we generalize our definition of a chain to join those measurements in a single chain. For example, a chain which contains only measurements with no quality cuts and both photodiodes active would double its length to be $2K$.

Naively, for a pair of simultaneous measurements in a chain, denoted by $k$ and $k'$ one would expect that $\phi_k = \phi_{k'}$. However, in our system this is not the case. Our measurement of $A_x + iA_y$ is also sensitive to the lock-in phase, which is different for the two photodiodes. In Eq. (S38), we have originally absorbed this phase in the unknown phase of $\bar{A}$, but now that the two measurements have the same $\bar{A}$ those phases have to be included explicitly. Moreover, our calibration data logged only the absolute value of $|A_x|$ and $|A_y|$ for each of the photodiodes, which introduces sign ambiguities and therefore phase ambiguities. We account for these effects in our model by assuming that $\phi_k - \phi_{k'}$ is uniformly distributed. The only exceptions to this model are pairs of measurements $k, k'$ where no Floquet field was active (and hence no lock-in detection). In this case, the two photodiodes measured the same definite orientation and therefore we use $\phi_k = \phi_{k'}$. Our likelihood then remains the same as Eq. (S57) except for a single modification; For every pair of simultaneous measurements $k, k'$ in a non-Floquet configuration we further multiply Eq. (S57) by $2\pi\delta(\phi_{k'} - \phi_k)$ inside the integral.

### 3. Full likelihood

To compute the likelihood numerically, we first approximate the phase integrals in Eq. (S57) analytically. For simplicity, in this section we will only treat chains with no non-Floquet measurements, so all the phases in the chain are statistically independent. The extension however to include such measurements is trivial. First, we note that

$$\mathbf{J}_{\text{tot}}^{\dagger} M_{\text{tot}}^{-1} \mathbf{J}_{\text{tot}}/2 = \sum_{k=1}^{K} \sum_{l=1}^{K} (|\mathbf{J}_k^{\dagger} M_{\text{tot}}^{-1} \mathbf{J}_l| \cos(\phi_l - \phi_k + \delta\phi_{lk}))/2, \tag{S61}$$

where

$$\delta\phi_{lk} = \text{Arg}(\mathbf{J}_k^{\dagger} M_{\text{tot}}^{-1} \mathbf{J}_l). \tag{S62}$$

Similar to Eq. S9, we now use an identity

$$e^{a\cos\phi} = \sum_{n=-\infty}^{\infty} I_n(a) e^{in\phi}, \tag{S63}$$

where here $I_n(a)$ denotes the modified Bessel function of the first kind, to rewrite the $\phi$s-dependent term in the likelihood,

$$e^{\mathbf{J}_{\text{tot}}^{\dagger} M_{\text{tot}}^{-1} \mathbf{J}_{\text{tot}}/2} = e^{\sum_k \mathbf{J}_k^{\dagger} M_{\text{tot}}^{-1} \mathbf{J}_k/2} \prod_{k=1}^{K-1} \prod_{l=k+1}^{K} \sum_{n_{lk}=-\infty}^{\infty} I_{n_{lk}} \left( \left| \mathbf{J}_k^{\dagger} M_{\text{tot}}^{-1} \mathbf{J}_l \right| \right) e^{in_{lk}(\phi_l - \phi_k + \delta\phi_{lk})}. \tag{S64}$$

Since $\int_0^{2\pi} e^{in\phi} d\phi$ is $2\pi$ if $n = 0$ and $0$ otherwise, we can now easily perform the angular integrals,

$$\left(\prod_{k=1}^{K} \int \frac{d\phi_k}{2\pi}\right) \prod_{k=1}^{K-1} \prod_{l=k+1}^{K} e^{in_{lk}(\phi_l - \phi_k)} = \prod_{k=1}^{K} \Delta\left(\sum_{l>k} n_{lk} - \sum_{k>l} n_{lk}\right), \tag{S65}$$

where $\Delta(x)$ is 1 if $x = 0$, and 0 otherwise. This gives K conditions on the $n_{lj}$s (and one can easily see that only $K - 1$ are independent).

Therefore, the final likelihood we used is yielded by substituting Eq. (S64) in Eq. (S57), and performing the $K\, d\phi_k$ integrals using Eq. (S65). This yields a function with only sums and products and no further integrals, so it can be computed directly. Numerically, we truncate the series in Eq. (S63) and sum a finite number of terms such that the $L_2$ norm of the difference between our sum and the original expression is at most 20%. Even so, and especially around $10 - 20$ Hz, the sum over all the different $n_{kl}$s had over $10^9$ terms in some cases, so that it was not feasible to carry out the full computation. In such cases, we narrowed the number of measurements down to only those where $m_{\rm DM} c^2 / h$ was closest to the resonance frequencies, until it became practical to carry out the computation.

Once the likelihood is computed, we use the Profile Likelihood Ratio (PLR) test to find the 95%C.L. bound on $g_{\rm a-Xe}$ [c.f. Eq. (S37)] for an ALP of mass $m_{\rm DM}$, by checking when

$$-2\log(L(g_{\rm a-Xe}, m_{\rm DM})/L(\hat{g}_{\rm a-Xe}, m_{\rm DM})) = 2.71, \tag{S66}$$

with $\hat{g}_{\rm a-Xe} > 0$ as the coupling that gives the maximal likelihood given the mass. When $\hat{g}_{\rm a-Xe} \neq 0$, it is possible that two $g_{\rm a-Xe}$ solve Eq. (S66) (one larger and one smaller than $\hat{g}_{\rm a-Xe}$), and we choose the solution which is larger than $\hat{g}_{\rm a-Xe}$. The value in the RHS, 2.71, is derived using the asymptotic so-called "half $\chi^2$" distribution [64].

For the quadratic type interactions, we define

$$g_{\rm q-Xe} \equiv \epsilon_N g_{\rm N-Quad}^2 + \epsilon_P g_{\rm P-Quad}^2, \tag{S67}$$

analogous to $g_{\rm a-Xe}$ and measured in units of squared inverse energy. As mentioned in the main text, there is a simple conversion formula between the two bounds, so no independent analysis was necessary. Instead, the bound can be derived from the following formula (c.f. Eqs. (1,3),

$$g_{\rm q-Xe,95\%}(m_{\rm DM}) = \frac{m_{\rm DM} g_{\rm a-Xe,95\%}(2m_{\rm DM})}{\sqrt{2\rho_{\rm DM}\hbar^3 c}}, \tag{S68}$$

where $g_{\rm q-Xe,95\%}(m_{\rm DM})$ is the 95%C.L. bound on $g_{\rm q-Xe}$, and computed for DM of mass $m_{\rm DM}$. $g_{\rm a-Xe,95\%}(2m_{\rm DM})$ is the 95%C.L. bound on $g_{\rm a-Xe}$, computed for DM of mass $2m_{\rm DM}$.

### E. Second stage analysis

We applied the analysis in S7 D in a blinded fashion, aside from a single minor post-unblinding change explained in S7 F, and obtained the constraints given in the main text. The statistical test enabled 95% CL exclusion of DM interactions, with the reported bounds in the main text. At several specific ALPs frequencies however, the likelihood test excluded our background-only hypothesis by the pre-ublinding designated threshold,

$$-2\log(L(g_{\rm a-Xe} = 0)/L(\hat{g}_{\rm a-Xe})) > 22.7, \tag{S69}$$

which is equivalent to $p_{\rm value} < 10^{-6}$ within the asymptotic limit. For the specific frequencies in which the test fails, we exclude the hypothesis of a white, flat noise in all these measurements. This scenario was expected in the design of the analysis: the likelihood model assumes that the noise model is *white*, whereas in practice, at several frequencies the noise has coherent magnetic peaks (as observed in control data that was taken before the experiment started). In this section, we use a second statistical test that assumes *magnetic coherent noise*, or a transient noise that appears only during a single measurement. We show that such a noise model does indeed explain the observed excess at all data points which were analyzed. Notably, this part of the analysis was conducted post-unblinding; We detail the changes and modifications with respect to the original analysis plan in subsection S7 F.

The first-stage analysis resulted with 52 bins whose spectral size is 1 mHz in the frequency domain that had at least one mass that failed the test we have postulated. This is a small fraction of the entire search data, which includes about one million such bins, indicating that most of our spectrum is consistent with the white-noise model. ALPs are postulated to be a non-transient object, and should be repeatedly measured at different instances, except for some statistical variation of their amplitude. The first action we applied was to run our original likelihood test again for each of those frequencies, but with the removal of a single measurement which had the largest amplitude at one of the frequencies within the bin. This action should remove measurements that experienced excess transient noise. Due to a lack of many measurements with hourly calibrations, we had less control over their backgrounds, and masses that were affected by such measurements were rerun only without these measurements, and without any additional measurement removed. As we had an ample sampling of the axial magnetic fields in the search (with respect to the NMR linewidth), the constraint data of ALPs is typically composed of at least 3-4 measurements with a resonance frequency within $\sim 2\Gamma_{\text{Xe}}$ of the central frequency. While the removal of the most peaked measurement is a clear introduction of bias, the remaining measurements are enough so that sensitivity is not entirely lost.

After re-running this test, 18 of the 52 bins passed the test of Eq. (S69), indicating that aside from a single measurement, their data was consistent with our white noise model. These 18 bins were centered around the following frequencies: 11.3862, 14.6715, 16.1699, 16.1724, 44.6417, 64.0148, 83.3353, 138.5963, 440.9693, 440.9706, 440.9721, 449.0672, 449.0691, 449.0714, 995.4164, 995.4176, 995.4194, 995.4218 Hz.

For the other masses within the other 34 bins, the presence of a strong, non-transient and semi-coherent signal excludes our white noise explanation, even with one of the measurements removed. We therefore devise a likelihood test in which the noise is assumed to be a coherent magnetic peak. ALPs signals are attenuated significantly away from the NMR resonance, since they couple solely to the xenon nuclei. Magnetic noise on the other hand, is coupled also to the rubidium spins, and therefore when it is set on the NMR resonance or away from it (within the magnetometer bandwidth), its amplitude changes moderately by $1/(1 + \xi)$ which is of order unity. To simplify the computational complexity of this particular test we ignore here the stochastic effects of the ALPs, and treat them as entirely coherent within each measurement, but uncorrelated between different measurements. We also ignored the temporal dependence of $\hat{x}(t), \hat{y}(t)$ (c.f. section S6 B), which is unimportant for the measurement time of the suspected 34 bins, shorter considerably than a day. Under these assumptions, for each ALP mass and a given continuous measurement, the matrix $\alpha_{mn,i}$ of Eq. (S40) becomes a single number $\alpha$. As the new test for the ALP mass $m_{\text{DM}}$ is constructed of N different continuous measurements, we denote the $\alpha$ that corresponds to the $j^{\text{th}}$ measurement by $\alpha_j$. Similarly, the matrices $\Sigma_N$ and $\Sigma_A$, and the vector $\mathbf{d}$, also become scalars, denoted by $\Sigma_{N,j}, \Sigma_{A,j}$, and $d_j$ respectively, for each of the $1 \leq j \leq N$ measurements. Furthermore, we only use our primary diode for this second analysis stage (as originally planned before unblinding) due to our secondary diode being more unstable, and thus more prone to transient noises.

Formally, we now consider the likelihood of the form

$$\mathcal{L} = \prod_{j=1}^{N} L_j(\alpha_j, d_j, \Sigma_{A,j}, \Sigma_{N,j}).$$

(S70)

using the functional form of $L_1$ of Eq. (S50) for each of the $L_j$s. However, in contrast to our first test, here $\Sigma_N$ is not predetermined by sidebands, and is rather assumed to be sourced by a magnetic peak, with a consistent amplitude,

$$\Sigma_{N,j} = A_B^2 \left( |A_{x,j} + iA_{y,j}| \left( 1 + \left| \frac{\xi_j}{i(f_{\text{Xe},j} - m_{\text{DM}}c^2/h)/\Gamma_{\text{Xe},j} + 1} \right| \right) \right)^2$$

(S71)

where $A_B$ is the unknown amplitude of the magnetic peak. We now apply the likelihood test and calculate the likelihood for the null hypothesis in which $g_{\text{a-Xe}} = 0$ but $A_B$ is a fit parameter, which maximizes $\mathcal{L}_0 \equiv \mathcal{L}(g_{\text{a-Xe}} = 0, A_B = \hat{A}_B)$ at $A_B = \hat{A}_B$. We then compare it to the model in which the ALP present and calculate two fitting parameters, $g_{\text{a-Xe}}$ and $A_B$ which maximize simultaneously $\mathcal{L}_1 \equiv \mathcal{L}(g_{\text{a-Xe}} = \hat{g}_{\text{a-Xe}}, A_B = \hat{A}_B)$ at $g_{\text{a-Xe}} = \hat{g}_{\text{a-Xe}}$ and $A_B = \hat{A}_B$. We then construct

$$\lambda = -2\log(\mathcal{L}_0/\mathcal{L}_1)$$

(S72)

and compute it for the different masses.

The value of the largest $\lambda$ in each of the 34 bins is given in Table S7 E. The $\lambda$s of all the masses in this test falls below our predetermined detection threshold of 22.7, showing they are consistent with the null hypothesis of no DM interactions. All values fall considerably below the threshold, except for the masses within the bin of $m_{\text{DM}}c^2/h = 704.2919$ Hz. While the resulting $\lambda = 22.4$ is marginally below the previously decided threshold, it is still rather high. We therefore ran the test again upon removal of a singular measurement and found that it drops to $\lambda = 0$, for all frequencies within the bin. We therefore interpret that at this frequency, we had a coherent magnetic noise that had transient variation in one of the experiments.

The 34 bins have the following spectral pattern. 21 bins were at most 0.1 Hz away from either 64.026 Hz (which is the refresh rate of some of the electrical components in the lab where the experiment was performed), or one of its harmonics. Of the remaining 13 bins, 10 were within 0.1 Hz of an harmonic of 60.04 Hz, and the last three bins with an excess had noise surrounding 31.62, 120.3, 140.14 Hz. This pattern is consistent with a few magnetic sources which caused the peaked coherent noise.

| frequency [Hz] | $\lambda_{\max}$ | frequency [Hz] | $\lambda_{\max}$ | frequency [Hz] | $\lambda_{\max}$ |
|---|---|---|---|---|---|
| 31.6173 | 0.0 | 60.1252 | 0.0 | 64.0265 | 0.0 |
| 64.0382 | 0.45 | 119.9827 | 0.0 | 119.9842 | 0.0 |
| 120.0117 | 0.0 | 120.013 | 0.0 | 120.0385 | 0.0 |
| 120.1156 | 0.0 | 120.3028 | 0.0 | 128.0528 | 0.0 |
| 128.054 | 0.0 | 140.1352 | 0.0 | 180.0578 | 0.0 |
| 192.0785 | 0.0 | 192.0806 | 0.0 | 240.0766 | 0.0 |
| 240.0777 | 0.0 | 256.1041 | 0.01 | 256.1058 | 0.0 |
| 256.1077 | 0.0 | 320.1315 | 0.0 | 320.1332 | 0.0 |
| 384.1593 | 0.26 | 448.1846 | 0.0 | 448.1862 | 0.0 |
| 704.2897 | 0.0 | 704.2908 | 0.0 | 704.2919 | 22.43, (0.0*) |
| 768.3151 | 3.01 | 768.317 | 1.1 | 768.3188 | 0.0 |
| 768.3201 | 0.0 | | | | |

TABLE S-I. **The results of the second analysis stage** - showing that suspicious frequencies which are inconsistent with white noise model are nonetheless consistent with coherent magnetic noise. Each frequency corresponds to the central frequency of a bin of width 1 mHz, for which the null hypothesis in the initial likelihood test, which assumes white-noise only, has failed. Here $\lambda_{\max}$ is the maximal twice log likelihood ratio (see Eq. (S72)) of any frequency that was inside the bin for a second likelihood test, now assuming a null hypothesis of coherent magnetic noise. All bins now pass our exclusion criteria and support the null analysis. Only a single bin, with frequency 704.2919 Hz only marginally passes the second stage of the analysis. However, this is due to the transient effect of a single measurement; removing that single measurement from a chain yields the value $\lambda = 0$ (denoted with an $^*$) for that frequency and supports the null hypothesis. ALPs generate non-transient signal and thus it is significantly more likely that a single measurement had an increased noise, rather than this would be explained by an ALP.

## F. Post Unblinding Changes

As previously mentioned, before the data-taking sessions began, control measurements of the spectrum were performed, both with and without an active pump beam, to measure the noise spectrum of the detector. These measurements showed that the noise was dominated by the photon-shot noise in most frequencies, and that when magnetic noise dominates, it is with sharp peaks. It is due to these control measurements, that it was believed that aside from several distinct frequencies with a large magnetic noise, the noise could be well described by photon shot noise. For technical reasons, masses nearby those that failed the test of Eq. (S69) would have also needed to be analyzed in the second stage of the analysis due to the participation of noisy points in their first stage noise estimation.

After unblinding, the observed 52 bins of noise were more numerous than the expected $\leq 10$ (though still a small fraction of all frequencies), and thus it was decided to not perform the full second stage of the analysis that was originally planned, and instead take a simpler approach. We therefore chose to remain with our original first-stage analysis for all constraints shown in this work. While the second stage of the analysis yields improved results for the frequencies in the 52 bins, as it was modified post-unblinding, we have opted to not use it in any of the constraints. The only minor necessary post-unblinding change that has affected our constraints was a change in how we estimate $\Sigma_N$. Masses that were near a magnetic peak, had used the magnetic peak for their estimation of the white noise, and were thus over-estimating the white noise, thus giving bounds which were much better than their true bound-setting ability. Originally, it was planned to rerun our secondary analysis for these points, but as that was not performed, to avoid showing bounds which were stronger than the ability of the experiment we changed our procedure to estimating $\Sigma_N$ to one that is much less sensitive to peaks. The change in the protocol was to change from taking the variance of the points, to computing the median in the norm squared of the $M$ points above, and the norm squared in the $M$ points below, and take the mean of these two norms, and use that as the estimate of the variance (up to an appropriate normalization constant of 2.2 as the median of the norm squared of a gaussianly distributed variable is 2.2 times smaller than its mean). This protocol gives a white noise estimate which is nearly identical to the previous one, unless it is computed in the presence of sharp peaks. Near sharp peaks, this protocol gives an estimate of the white noise which would be insensitive to the sharp peaks, as wanted. For the sake of consistency, we have applied this to all masses, though we again emphasize that it had a negligible difference to all points aside from those that neighbor a magnetic peak.

We now move to discuss the changes for the post-unblinding second stage of the analysis. The performed second stage of the analysis is fully described in subsection S7 E. Originally, it was not planned to remove any measurement. Our original control measurements which were used to design our analysis procedure were taken in a period significantly shorter than 5 months, which meant we missed the existence of transient noises that may appear once in many days. Therefore, the examination of the removal of measurements from the analysis to remove transient noises was not assumed to be necessary pre-unblinding.

After the rerunning of the first analysis while removing a single measurement for each of the masses that failed the test of Eq. (S69), the remainder of the analysis that was performed was the exact analysis that was originally planned, albeit it included several simplifications. Originally, it was planned to use the full stochastic modeling of the ALPs, as described for the first stage of the analysis. However, we have opted to approximate it as a monochromatic signal. As all measurements used for this second analysis were at most a third of the total coherence time, $\Sigma_A$ was nearly concentrated within a central frequency regardless. The neglecting of ALP correlation between different measurements is somewhat more meaningful, but since the majority sequentially ordered measurements were spaced by $\sim 5\Gamma_{Xe}$, measurements that were highly sensitive to the same ALP mass, were rarely within a single coherence time from each other, rendering this effect to be small.

After these changes, the analysis had one relevant frequency from each measurement. Due to that, adding the white noise of the photon shot noise was entirely irrelevant (as in that frequency, the magnetic noise dominated), and we have thus removed it from consideration of the noise model. Regardless of the exact performed analysis, we emphasize that the control measurements show that most of the 34 bins are near the harmonics of only a few very noisy frequencies, and are extremely noisy even far from resonance. In the future, we plan to avoid such post-unblinding changes by having more robust control measurements, as well as to introduce the possibility of up to few hours-long transient noise in our quality cuts, which disappears at repeated measurements.

## S8. MODEL-DEPENDENT BOUNDS ON COUPLING OF DM WITH PROTONS

In this section, we present model-dependent constraints on the coupling of DM with protons. As discussed in the main text, these constraints are less-reliable due to an unbounded uncertainty of the exact proton fraction $\epsilon_P$ within the nuclear spin of $^{129}$Xe. Here we consider 4 proton fractions adapted from different models in Ref. [43], and are computed using four different models for the nuclear interactions. These correspond to the 1.8/2.0 (EM) model, the N$^4$LO+3N$_{lnl}$ model, the Large Scale Shell Model (LSSM), and their newest $\Delta$ NNLO$_{GO}$(394) model, which yield the proton fractions $\epsilon_P = 0.0036, -0.0084, 0.020, -0.0308$ respectively. Negative contribution corresponds to anti-alignment of the proton with respect to the total spin of the nucleus, and experimentally corresponds to a precession in the counter direction. Model-dependent bounds on ALP-proton couplings and on quadratic coupling of DM to protons based on these values are presented in Figs S6,S7 respectively.

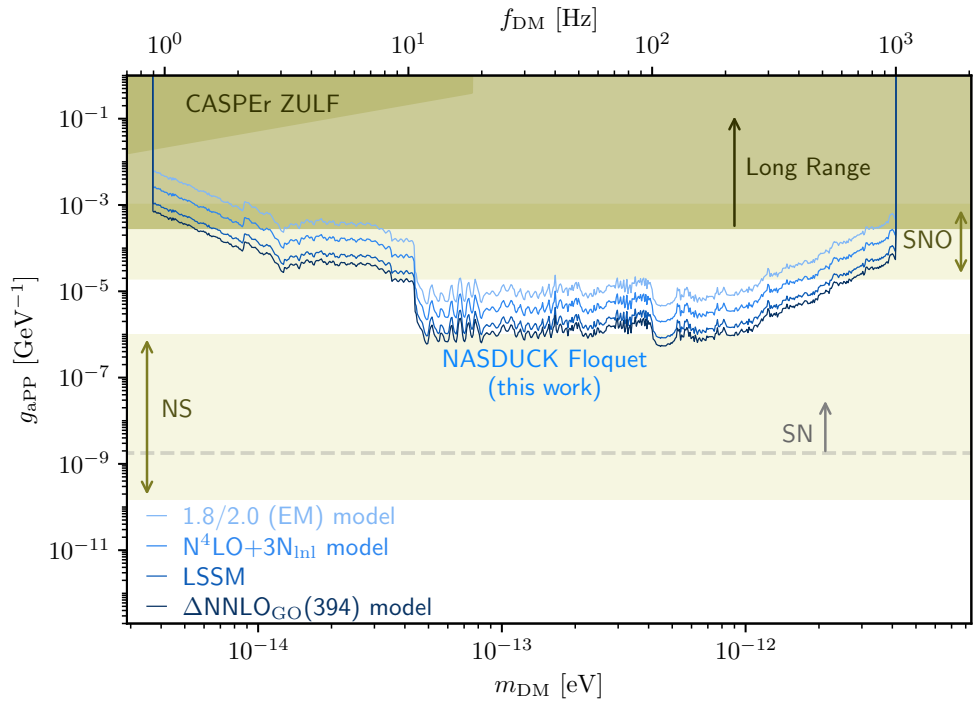

FIG. S6. **Model-dependent constraints on the ALP-proton couplings**. The bounds are derived using the Floquet quantum detector by the NASDUCK collaboration in this work. All presented constraints calculated in this paper correspond to 95% confidence level of the bound. The precise tabulated bounds can be found in [48], though the **blue shaded solid lines** show a binned average of the bound. The **olive-green** regions show other terrestrial constraints, including the CASPEr ZULF experiments [19,49] (with $\epsilon_P = 1$), and Long range constraints on ALP-proton couplings. In **beige**, the agreed astrophysical excluded region from solar ALPs unobserved in the Solar Neutrino Observatory (SNO) [51] and from neutron stars cooling [52,66,67,68] is shown. The region above the **gray dashed line** is excluded by supernova (SN) cooling considerations and neutrino flux measurements [4,53,54]. We stress that the SN cooling constraint strongly relies on the unknown collapse mechanism and hence the limits should be taken with a grain of salt [55]. Similarly, our own bounds also rely on the unknown proton fraction of the $^{129}$Xe nuclear spin, and they should be taken with a grain of salt as well.

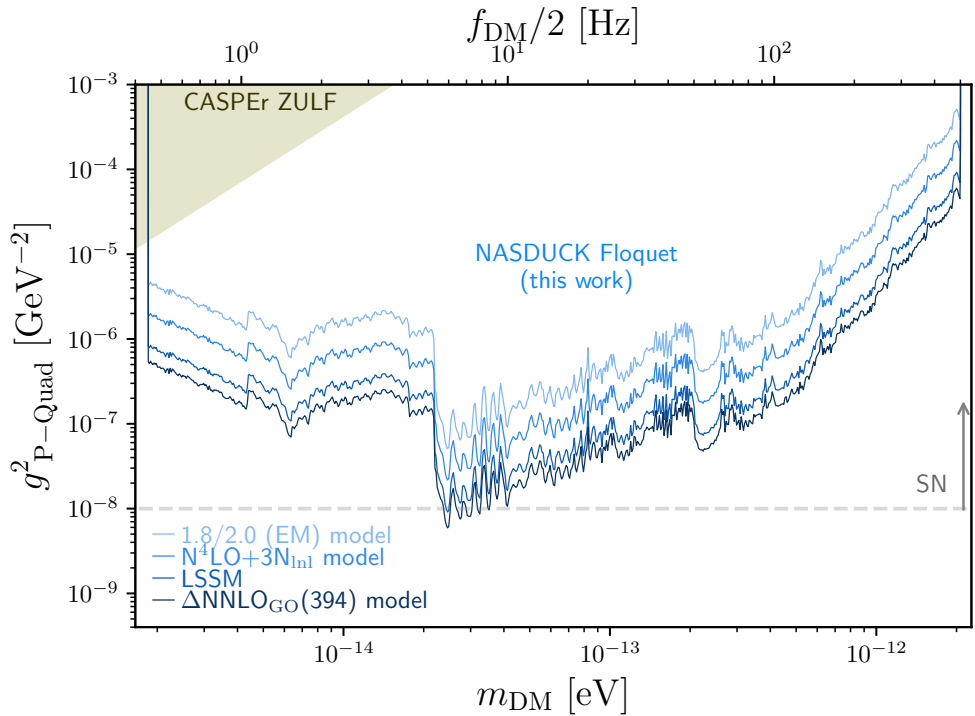

FIG. S7. **Model dependent constraints on nuclei-DM couplings of quadratic type**. The bounds are given as a function of the ALP mass, derived using the Floquet quantum detector by the NASDUCK collaboration in this work. All presented constraints calculated in this paper correspond to 95% confidence level of the bound. The precise tabulated bounds can be found in [48], though the **blue shaded solid lines** show a binned average of the bound. The **olive-green** regions show the CASPEr ZULF bound [12,49], for $\epsilon_P = 1$. The region above the **dashed gray line** is excluded by supernova (SN) cooling considerations [13]. We stress that the SN cooling constraint strongly relies on the unknown collapse mechanism and hence the limits should be taken with a grain of salt [55]. In addition, since our own bounds also rely on the unknown proton fraction of the $^{129}$Xe nuclear spin, they should also be taken with a grain of salt.

Notably, the different models give different values of $\epsilon_P$, whose sign is inconsistent, and are therefore consistent with a vanishing $\epsilon_P$ within the current theoretical uncertainty. Hence, the bounds presented in Figs S6,S7 should be taken with a grain of salt. Nonetheless, in the future, should the theoretical uncertainties on the proton fraction of the $^{129}$Xe spin be reduced, our bounds can be easily recomputed and updated by a simply normalizing $\epsilon_P$ using the online available data [48].